\documentclass[useAMS,usenatbib]{mn2e}

\usepackage{times}  % to get nice fonts for PDF
\usepackage{listings}
\usepackage{graphics}
\usepackage{subfigure}
\usepackage{graphicx}
\usepackage{graphicx,xspace}
\usepackage{amssymb}
\usepackage{amsmath}
\usepackage{hyperref}
\usepackage{aas_macros}
\usepackage{lscape}
\usepackage{rotating}
\usepackage{placeins}
\usepackage{natbib}
\usepackage{color}
\usepackage[table]{xcolor}
     
\newcommand{\galform}{{\sc{galform}}\xspace}

\newcommand{\eagle}{{\sc{eagle}}\xspace}
\newcommand{\gadget}{{\sc{gadget-3}}\xspace}
\newcommand{\Lgal}{{\sc{L-Galaxies}}\xspace}
\newcommand{\subfind}{{\sc{subfind}}\xspace}
\newcommand{\things}{{\sc{things}}\xspace}
\newcommand{\owls}{{\sc{owls}}\xspace}
\newcommand{\dhalo}{{\sc{dhalo}}\xspace}

\title[Comparing galaxy formation in semi-analytic models and hydrodynamical simulations]
{Comparing galaxy formation in semi-analytic models and hydrodynamical simulations}
\author[P. D. Mitchell et al.]{
\newauthor Peter D. Mitchell\thanks{\rm E-mail: peter.mitchell@univ-lyon1.fr}$^{1}$,
Cedric G. Lacey$^{2}$,
Claudia D. P. Lagos$^{3}$,
Carlos S. Frenk$^{2}$,
\newauthor Richard G. Bower$^{2}$,
Shaun Cole$^{2}$,
John C. Helly$^{2}$,
Matthieu Schaller$^{2}$,
\newauthor Violeta Gonzalez-Perez$^{4}$,
Tom Theuns$^{2,5}$
\\
$^{1}$Univ Lyon, Univ Lyon1, Ens de Lyon, CNRS, Centre de Recherche Astrophysique de Lyon UMR5574, F-69230, Saint-Genis-Laval, France\\
$^{2}$Institute for Computational Cosmology, Department of Physics, University of Durham, South Road, Durham, DH1 3LE, UK\\
$^{3}$International Centre for Radio Astronomy Research, 7 Fairway, Crawley, 6009, Perth, WA, Australia\\
$^{4}$Institute of Cosmology and Gravitation, Portsmouth University, Dennis Sciama Building, Burnaby Road, Portsmouth PO1 3FX, UK\\
$^{5}$Department of Physics, University of Antwerp, Groenenborgerlaan 171, B-2020 Antwerpen, Belgium
}
\begin{document}
\date{\today}
\pagerange{\pageref{firstpage}--\pageref{lastpage}} \pubyear{2017}
\maketitle
\label{firstpage}

\begin{abstract}
It is now possible for hydrodynamical simulations to reproduce a representative galaxy population.
Accordingly, it is timely to assess critically some of the assumptions of traditional semi-analytic
galaxy formation models. We use the \eagle simulations to assess assumptions built into the
\galform semi-analytic model, focussing on those relating to baryon cycling, angular momentum
and feedback.
We show that the assumption in \galform that newly formed stars have the same specific angular momentum
as the total disc leads to a significant overestimate of the total stellar specific angular momentum of
disc galaxies. In \eagle, stars form preferentially out of low specific angular momentum gas
in the interstellar medium (ISM) due to the assumed gas density threshold for stars to form, leading to more realistic galaxy sizes.
We find that stellar mass assembly is similar between \galform and \eagle but that the evolution
of gas properties is different, with various indications that the rate of baryon cycling in \eagle is
slower than is assumed in \galform. 
Finally, by matching individual galaxies between \eagle and \galform, we find that an artificial dependence
of AGN feedback and gas infall rates on halo mass doubling events in \galform drives most
of the scatter in stellar mass between individual objects. Put together our results suggest that 
the \galform semi-analytic model can be significantly improved in light of recent advances.
\end{abstract}


\section{Introduction}

Semi-analytic galaxy formation models are established tools for connecting the
predicted hierarchical growth of dark matter haloes to the observed properties
of the galaxy population \cite[e.g.][]{Cole00,Somerville08,Guo11}. Unlike empirical 
abundance matching \cite[e.g.][]{Conroy06,Moster10}
or halo occupation distribution models \cite[e.g.][]{Berlind02}, semi-analytic models 
employ a forward-modelling approach and are constructed such that they contain as 
much as possible of the baryonic physics that is thought to be relevant to galaxy 
evolution, albeit at a simplified, macroscopic level. The simplified, macroscopic 
nature of semi-analytic models means that they are computationally inexpensive to evaluate. 
Compared to hydrodynamical simulations, this lack of computational expense meant
that until recently it was uniquely possible for semi-analytic models to predict 
realistic galaxy populations \cite[e.g.][]{Bower06,Croton06,Henriques13}.

Recently, advances in computational resources combined with improvements in
the uncertain modelling of feedback have allowed hydrodynamical simulations
to predict galaxy populations which reproduce observations at an
equivalent level to semi-analytic models for representative 
volumes \cite[][]{Vogelsberger14,Schaye15,Dubois14,Dave16}. It is timely therefore
to review the underling assumptions which underpin semi-analytic models
and assess their validity against state-of-the-art hydrodynamical simulations.

As in semi-analytic models, hydrodynamical simulations are forced to 
implement uncertain subgrid modelling to approximate the effect of massive 
stars and black holes on galaxy evolution. This means that, for example,
the dynamics of outflowing gas in these simulations are not necessarily
realistic (irrespective of whether a realistic galaxy population is
produced). Importantly however, the dynamics of outflows are tracked
self-consistently in hydrodynamical simulations. Furthermore, simulations 
do not need to make any assumptions regarding angular momentum conservation 
of the various baryonic components of galaxies and their surrounding gas 
flows. In semi-analytic models, both
these aspects of galaxy evolution are among the most important for predicting
galaxy properties and yet are also among the most uncertain 
\cite[][]{Henriques13,Mitchell14,Hirschmann16}. Arguably therefore, the parametrisations
of these physical processes that are implemented  in semi-analytic models
should be capable of reproducing (with an appropriate choice of model 
parameters) the behaviour predicted by hydrodynamical simulations.
Here, we begin to address this topic by comparing the properties
of galaxies between the established semi-analytic model, \galform \cite[][]{Cole00,Lacey16},
and the \eagle simulation project, a state-of-the-art suite of calibrated
hydrodynamical simulations \cite[][]{Schaye15,Crain15}.

A number of other authors have presented comparisons of results from hydrodynamical simulations to semi-analytic
models, albeit without access to hydrodynamical simulations that predict realistic
galaxy properties for representative volumes. Some studies have focused specifically
on cooling and gas infall onto galaxies, finding varying levels of agreement 
\cite[][]{Yoshida02,Helly03,Monaco14}. \cite{Saro10} analysed a single, massive
cluster, finding significant differences in the manner with which tidal stripping
acts between a semi-analytic model and a hydrodynamical simulation.
\cite{Stringer10} analysed a single disc galaxy, finding it was possible to roughly
reproduce a hydrodynamical simulation with an explicitly calibrated semi-analytic model.
\cite{Cattaneo07} and \cite{Hirschmann12} analysed larger samples of galaxies, both
finding broad agreement in stellar and baryonic masses but significant differences when 
analysed in detail. In particular, \cite{Hirschmann12} reported large differences in
star formation efficiency stemming from local versus global implementations of
a Kennicutt star formation law.

This study follows from \cite{Guo16}, who compared \galform, the similar
\Lgal model \cite[][]{Guo11} and \eagle. They focused on global
predictions for the galaxy population (stellar mass functions, star formation 
rates (SFRs), passive fractions, mass-metallicity relations, mass-size relations). 
They showed that stellar mass functions and passive fractions were broadly similar
between the models (provided gradual ram pressure stripping of hot gas from 
satellites was implemented in \galform). However, they also showed that
predictions for galaxy sizes differed significantly and that mass-metallicity 
relations are significantly steeper in \galform than in the reference \eagle model.
In both cases, the predictions from \eagle are in significantly better agreement
with observations.

While these disagreements between the models are highly suggestive of differing
baryon cycling (because of discrepant metallicities) and angular momentum evolution \cite[because of discrepant sizes, see also][]{Stevens17}, 
from a global comparison it  is not clear exactly how these differences arise. Here,
we compare the \galform and \eagle models in more detail and attempt 
to isolate as far as possible distinct physical processes, focussing on those
which we expect may not be modelled realistically in semi-analytic models.
We also match individual galaxies by matching the
haloes between the dark-matter-only version of \eagle and the full hydrodynamical
simulation. This allows us to assess the difference in stellar mass between individual
galaxies. 

The results and methodology presented here will, in turn, underpin a future study where we plan
to perform the most direct level of comparison possible between \galform and \eagle.
Namely, to directly measure all of the mass, metal and angular momentum exchanges
between different discrete baryonic reservoirs in \eagle and compare with the corresponding
quantities in \galform. As such, we consider here how to compartmentalize
baryons in \eagle between the corresponding discrete components that are tracked
in semi-analytic models. In particular, we carefully consider how to separate the 
interstellar medium (ISM) from more diffuse halo gas in the circumgalactic medium 
(CGM) in \eagle on physical grounds.

The layout of this paper is as follows. We introduce the \eagle simulations,
the \galform semi-analytic model and describe our analysis methodology in Section~\ref{model_section}.
We present
a first comparison of the models by analysing stellar masses in 
Section~\ref{sm_section}. We compare star formation thresholds and efficiencies
as well as the angular momentum of star-forming gas in Section~\ref{sf_section}.
We discuss feedback from supernovae (SNe) and active-galactic nuclei (AGN) in
Section~\ref{fb_section} and the resulting baryon cycle in Section~\ref{cycle_section}.
We discuss the consequences of qualitative differences between gas infall rates 
onto galaxies in the two models in Section~\ref{infall_section}. Finally, we summarise
our main results in Section~\ref{summary_section}.

Throughout this paper we denote the units of distances in proper kiloparsecs as $\mathrm{pkpc}$
and comoving kiloparsecs as $\mathrm{ckpc}$.
Also throughout, $\log$ refers to the base $10$ logarithm and $\ln$ refers to the 
natural logarithm.

\section{Modelling galaxy formation}
\label{model_section}

To facilitate a direct comparison of the \eagle simulations and the \galform model, we follow \cite{Guo16}
by running \galform on a dark-matter-only version of the reference \eagle simulation run with a 
$100^3 \, \mathrm{Mpc^3}$ box \cite[L100N1504 in the convention introduced by][]{Schaye15}.
As described by \cite{Guo16}, both simulations where performed with the same cosmological parameters 
taken from \cite{Planck14}, and with the same initial conditions, following the method of \cite{Jenkins10}.

\subsection{\eagle}

The \eagle simulations are a suite of hydrodynamical simulations of the formation 
and evolution of galaxies within the context of the $\Lambda$CDM cosmological model. 
Performed using a modified version of the \gadget code 
\cite[last presented in][]{Springel05}, they include a state-of-the-art implementation of smoothed particle hydrodynamics
\cite[SPH, Dalla Vecchia in prep,][]{Schaller15a}. They also include a set of subgrid models that account for the physics of
photo-heating/ionization from an evolving, uniform background radiation field, radiative cooling from metal 
lines and atomic processes, star formation, stellar and supermassive black hole evolution 
and feedback. Subgrid models are included to compensate for the limited resolution of cosmological simulations and
the prohibitive computational cost of performing detailed on-the-fly radiative transfer. 
A detailed overview of these subgrid models can be found in \cite{Schaye15} and a concise 
overview tailored to the topic of comparison with semi-analytic models can be found in 
\cite{Guo16}.

\subsection{\galform}

\galform is a continually updated semi-analytic galaxy formation model, first introduced in
\cite{Cole00}, which itself was an evolution from earlier models \cite[e.g.][]{Lacey93,Cole94}.
Salient updates subsequent to \cite{Cole00} include the inclusion of AGN feedback \cite[][]{Bower06},
the addition of gradual ram-pressure stripping in satellites \cite[][]{Font08} and a decomposition 
of the ISM into neutral atomic and molecular hydrogen components \cite[][]{Lagos11a}. The most
recent branches of the model can roughly be divided between a model with a universal stellar
IMF \cite[][]{GonzalezPerez14, Guo16},
and a model that also implements a non-standard IMF in nuclear starbursts \cite[][]{Lacey16}.

\cite{Guo16} introduced a version of the universal IMF model that was explicitly tuned for
the \eagle DM-only simulation using $200$ simulation outputs. For this study, we use an 
updated version of this model which is very similar. The updates were made to ensure that properties
of individual galaxies do not depend on random numbers, as discussed below. This is a desirable 
step for comparing with the \eagle simulations on an object-by-object basis.
Otherwise, the model parameters are the same as \cite{Guo16}, with the exception of slight changes
to two parameters which control the efficiencies of SNe and AGN feedback\footnote{Specifically,
we change the normalisation of the SNe feedback mass loading factor, $V_{\mathrm{SN}}$ from 
$425 \, \mathrm{km \, s^{-1}}$ to $380 \, \mathrm{km \, s^{-1}}$ and the threshold for 
AGN feedback, $\alpha_{\mathrm{cool}}$, from $0.52$ to $0.8$. See \cite{Lacey16}
for the definition of these model parameters.}. These changes were made to 
approximately restore predictions for the local stellar mass function presented in \cite{Guo16} 
after an error in the calculation of halo concentrations introduced in \cite{Guo16} was corrected 
\footnote{Specifically, an incorrect tabulated power spectrum file was used to calculate halo
concentrations. This error does not affect
any of the conclusions of that study.}. The calibration of the reference model used here 
hence follows \cite{Guo16} and is described in Section~\ref{calibration}.

To remove a dependency of individual galaxy properties on random numbers, we make two changes
with respect to \cite{Guo16}. The first is that we now measure halo spin 
parameters from the \eagle dark-matter-only simulation instead of sampling from a probability
distribution function, as introduced in \cite{Cole00}. The second is that we now track
the orbits of satellites measured in the dark-matter-only simulation. Once satellites
can no longer be identified in the simulation, a self-consistent dynamical friction merging 
timescale, $T_{\mathrm{df}}$, is then computed as

\begin{equation}
T_{\mathrm{df}} = \left(\frac{R_{\mathrm{c}}}{R_{\mathrm{H}}}\right)^{1.8} \, \left(\frac{J}{J_{\mathrm{c}}}\right)^{0.85} \, \frac{\tau_{\mathrm{dyn}}}{2 B(1) \ln(\Lambda)} \left(\frac{M_{\mathrm{H}}}{M_{\mathrm{S}}}\right), 
\label{dfriction}
\end{equation}

\noindent where $R_{\mathrm{H}}$ is the halo virial radius, $R_{\mathrm{c}}$
is the radius of a circular orbit with the same energy as the 
actual orbit, $J/J_{\mathrm{c}}$ is the ratio of the angular momentum of
the actual orbit to the angular momentum of a circular orbit
with the same energy, $\tau_{\mathrm{dyn}}$ is the halo dynamical
time\footnote{Defining the halo dynamical time as $\tau_{\mathrm{dyn}} = R_{\mathrm{H}} / V_{\mathrm{H}}$,
where $R_{\mathrm{H}}$ is the halo virial radius and $V_{\mathrm{H}}$ is the
halo circular velocity at the virial radius.}, $M_{\mathrm{H}}$ is the
host halo mass, $M_{\mathrm{S}}$ is the mass of the satellite subhalo,
$\ln(\Lambda)$ is the Coulomb logarithm (taken to be $\ln(\Lambda)=\ln(M_{\mathrm{H}}/M_{\mathrm{S}})$)
and $B(x) = \mathrm{erf}(x) - 2x/\sqrt{\pi} \, \exp(-x^2)$. 
The full details of this new merging scheme are given in \cite{Simha16}\footnote{
Note that we do not use the tidal disruption model described in \cite{Simha16}.}.

\subsection{Structure and assumptions that underpin semi-analytic galaxy formation models}
\label{GalformOverview}

In semi-analytic galaxy formation models such as \galform, the initial presupposition is that
baryons trace the accretion of dark matter mass and angular momentum onto collapsed
dark matter haloes, that the baryons that have been accreted onto dark matter haloes can be 
compartmentalized into a few discrete components, and that these components can be 
adequately characterised by a handful of quantities.  The hierarchy of galaxy formation is 
accounted for by including each subhalo as a distinct entity such that the evolution 
of satellite galaxies is tracked within parent haloes. 

The discrete baryonic components typically tracked in a modern semi-analytic galaxy formation
model consist of a galaxy disc, a galaxy bulge/spheroid, a diffuse gas halo and a reservoir 
of gas that has been ejected from the galaxy by feedback. The quantities tracked for each 
of these components typically include the total mass, the magnitude of the angular momentum, the metal content 
and a set of scale lengths that specify the spatial distribution, assuming idealised density profiles. 
These quantities are evolved enforcing mass conservation and (typically) total angular momentum 
conservation. Galaxy formation is expected to be a highly dissipative process and so
energy conservation is not usually explicitly tracked \cite[although see][]{Monaco07}. 
However, individual physical processes do often contain energetic considerations, for example
in the computation of a radiative cooling timescale for hot diffuse halo gas.

A simplified, linearised version of the mass conservation equations for a central subhalo in
\galform is

%\begin{equation}
\begingroup
\renewcommand*{\arraystretch}{1.5}
%\begin{align*}
\begin{multline}
\begin{bmatrix}
  \dot{M}_{\mathrm{diffuse}} \\
  \dot{M}_{\mathrm{ISM}} \\
  \dot{M}_{\mathrm{ejected}} \\
  \dot{M}_{\mathrm{\star}} \\
\end{bmatrix}
= 
\begin{bmatrix}
  f_{\mathrm{B}} \dot{M}_{\mathrm{H}}\\
  0\\
  0\\
  0\\
\end{bmatrix}\\
+
\begin{bmatrix}
  -\frac{1}{\tau_{\mathrm{infall}}} & 0 & \frac{1}{\tau_{\mathrm{ret}}} & 0 \\ 
  \frac{1}{\tau_{\mathrm{infall}}} & -\frac{(1-R+\beta_{\mathrm{ml}})}{\tau_\star} & 0 & 0 \\
  0 & \frac{\beta_{\mathrm{ml}}}{\tau_\star} & -\frac{1}{\tau_{\mathrm{ret}}} & 0 \\
  0 & 0 & \frac{(1-R)}{\tau_\star} & 0 \\
\end{bmatrix}
\begin{bmatrix}
  M_{\mathrm{diffuse}} \\
  M_{\mathrm{ISM}} \\
  M_{\mathrm{ejected}} \\
  M_{\mathrm{\star}} \\
\end{bmatrix}
\label{ODE_MATRIX}
\end{multline}
%\end{align*}
\endgroup
%\end{equation}
\noindent where $M_{\mathrm{diffuse}}$ is the mass in a diffuse gas halo,
$M_{\mathrm{ISM}}$ is the mass in the interstellar medium, $M_{\mathrm{ejected}}$
is the mass in a reservoir of gas ejected from the galaxy by feedback, $M_\star$
is the mass in stars and $M_{\mathrm{H}}$ is the total halo mass. 
$f_{\mathrm{B}}$ is the cosmic baryon fraction, $\tau_{\mathrm{infall}}$
is the timescale for halo gas to infall onto a disc, $\tau_\star$ is the
disc star formation timescale, $R$ is the mass fraction returned from
stars to the ISM through stellar mass loss, $\beta_{\mathrm{ml}}$ is
the efficiency of SNe feedback\footnote{Specifically $\beta_{\mathrm{ml}}$
is the mass loading factor, defined as the ratio of the mass outflow
rate from galaxies to the star formation rate.} and $\tau_{\mathrm{ret}}$
is the return timescale for gas ejected by feedback.
Alongside Eqn~\ref{ODE_MATRIX}, there is also a corresponding set of equations for metal mass and angular momentum.
Here, we have neglected the distinction between the disc and bulge/spheroid
for simplicity and we have written the star formation law as being linear in the ISM gas
mass \cite[which is not the case for \galform models following][]{Lagos11a}.

The source term in Eqn~\ref{ODE_MATRIX} is the halo accretion rate, $\dot{M}_{\mathrm{H}}$,
scaled by the cosmic baryon fraction, $f_{\mathrm{B}}$, to give the baryonic accretion rate.
The stellar mass reservoir, $M_\star$, acts as a sink term because stellar recycling is implemented with 
the instantaneous recycling approximation \cite[][]{Cole00}. The strongly coupled nature of galaxy formation 
is encoded in the off-diagonal terms of Eqn~\ref{ODE_MATRIX}. 
Several of these terms ($\tau_{\mathrm{infall}}, \tau_\star$ and $\beta_{\mathrm{ml}}$)
depend in a non-linear fashion on different combinations of the halo density profile, 
halo mass accretion history, halo spin, disc angular momentum and diffuse halo metal content, 
such that the coupling between various aspects of the model is implicitly tighter than is 
shown explicitly in Eqn~\ref{ODE_MATRIX}.

The terms that appear in the central matrix of Eqn~\ref{ODE_MATRIX} represent 
distinct physical processes, some of which are analogous to the inclusion of 
the subgrid models included in \eagle. These include star formation ($\tau_\star$), 
stellar recycling ($R$) and the energy injection from stellar feedback 
($\beta_{\mathrm{ml}}$). Other processes are not modelled by subgrid models in \eagle and emerge 
naturally within the hydrodynamical simulation (although these will still
be affected by uncertain subgrid modelling). These include gas infall 
from a diffuse halo (characterised by $\tau_{\mathrm{infall}}$) and  
reincorporation from a reservoir of gas ejected from the galaxy by feedback
back into the diffuse halo ($\tau_{\mathrm{ret}}$).

Additional physical processes included in \galform but not shown in Eqn.~\ref{ODE_MATRIX}
include metal enrichment of the ISM, the growth of central SMBHs and the resulting AGN 
feedback, ram-pressure stripping of satellite galaxies, spheroid formation through
galaxy mergers and disc instabilities, nuclear starbursts and the suppression of
gas accretion onto small haloes by the UV background after reionization \cite[for a complete overview see][]{Lacey16}.

\begin{table*}
\begin{center}
\fontsize{9}{11}\selectfont
\begin{tabular}{|c|p{6.6cm}|p{6.6cm}|}
  \hline
  & \galform & \eagle \\
  \hline
  Star formation threshold & Disc star formation traces molecular hydrogen. Molecular hydrogen fraction depends on mid-plane gas pressure, which in turn depends on disc gas and stellar mass, and disc size.  & Star formation occurs in gas above a local density and metallicity dependent threshold. \\
  \hline
  Star formation law & Disc star formation rate is linear in molecular gas mass. & Star formation rate scales with local gas pressure, analogous to a Kennicutt-Schmidt law.\\
  \hline
  Disc angular momentum & Disc angular momentum is calculated assuming infalling gas conserves angular momentum. Gas and stars in the disc have equal specific angular momentum. & Angular momentum is computed locally following gravity and hydrodynamics.\\
  \hline
  Stellar feedback & Gas is ejected from galaxies (and haloes) as stars are formed, with an efficiency scaling with galaxy circular velocity. & Thermal energy is injected into the ISM around young star particles. The average energy injection scales with local gas density and metallicity. Gas is always heated by $\Delta T = 10^{7.5} \, \mathrm{K}$. \\
  \hline
  Black hole growth & SMBHs grow during galaxy mergers and disc instabilities. & SMBHs grow from surrounding ISM with an Eddington-limited Bondi accretion rate. \\
  \hline
  AGN feedback & SMBHs truncate gas infall onto the disc if the surrounding halo is quasi-hydrostatic and the SMBH injects enough energy to offset the radiative cooling rate. & Accreting SMBHs inject thermal energy into the surrounding ISM with a fixed average efficiency. Gas is always heated by $\Delta T = 10^{8.5} \, \mathrm{K}$. \\
  \hline
  Gas return & Gas ejected from haloes by stellar feedback returns to the halo over a halo dynamical time. & Gas return depends on particle trajectories which follow gravity and hydrodynamics. These trajectories are (presumably) sensitive to stellar and AGN feedback parameters, including the heating temperature. \\
  \hline
  Gas infall & Gas infall onto galaxy discs is either limited by gravitational freefall or radiative cooling timescales, depending on halo gas and dark matter density profiles, halo gas metallicity, and the time elapsed since halo mass-doubling events. & Gas infall is computed locally following gravity and hydrodynamics.\\
  \hline
\end{tabular}
\caption{Summary of the modelling of physical processes which are relevant to the results presented in this study,
sorted by the order in which these processes are discussed.}
\label{model_table}
\end{center}
\end{table*}

Table~\ref{model_table} presents a brief summary of the relevant physical processes included in \galform
and \eagle, sorted by the order in which they discussed in the following sections. More details for 
these physical processes are given in Appendix~\ref{ap:model_details}.

\subsection{Subhalo identification \& merger trees}
\label{subfind_section}

In both \galform and \eagle, haloes are identified first as groups using a friends-of-friends (FoF) algorithm,
adopting a dimensionless linking length of $b=0.2$ \cite[][]{Davis85}. FoF groups are split
into subhaloes of bound particles using the \subfind algorithm \cite[][]{Springel01,Dolag09}. For \eagle, 
galaxies are then defined as the baryonic particles bound to a given subhalo. For each FoF group, the subhalo 
containing the particle with the lowest value of the 
gravitational potential is defined as hosting the central galaxy and other subhaloes are defined as hosting 
satellite galaxies. Galaxy centres are based on the position of the particle with the lowest gravitational
potential.

In \galform, haloes are identified in the same way using the dark-matter-only version
of the reference $L100N1504$ \eagle simulation. Unlike in \eagle, groups of subhaloes are 
then grouped together by the \dhalo algorithm presented by \cite{Jiang14}.
This algorithm sets the distinction between central and satellite galaxies using 
information from the progenitors of a given subhalo.
In detail, subhaloes are flagged as satellites for the first time when they first enter within twice the half-mass radius 
of a more massive subhalo and if they have lost at least $25 \%$ of their past maximum mass 
(see Appendix A3 in Jiang et al. 2014). Once a subhalo is identified as a satellite, this status is then
preserved for all its descendants for which it is considered the main progenitor. This leads to situations
where galaxies are still considered to be satellites even if they have escaped outside the virial radius of a 
parent host halo and out into the field \cite[see][for a discussion of the importance of this choice and a comparison
with the implementation in the \Lgal model]{Guo16}. 

We define halo masses in \eagle (only for central subhaloes) as $M_{\mathrm{200}}$, the total mass 
enclosed within a radius within which the mean internal density is $200$ times the critical 
density of the Universe. Internally within \galform, halo masses are defined simply as the sum of
the mass of each subhalo associated with a given \dhalo (denoted as $M_{\mathrm{DH}}$). The masses of each subhalo are defined
simply as the sum of the particles considered gravitationally bound by \subfind to that subhalo.
We have also measured $M_{\mathrm{200}}$ for central galaxies from the \eagle dark-matter-only 
simulation and we use these masses for \galform galaxies when comparing galaxy properties at a given 
halo mass to \eagle. 
The differences between these various halo masses are shown in Appendix~\ref{ap:halo_mass}.
Hereafter, we denote $M_{\mathrm{H}}$ as referring to $M_{\mathrm{200}}$ measured from the 
hydrodynamical simulations for \eagle galaxies and $M_{\mathrm{200}}$ measured from the 
dark-matter-only simulation for \galform galaxies.

As well as the merger trees used in \galform, we also construct merger trees for the reference 
$L100N1504$ \eagle hydrodynamical simulation using the same subhalo merger tree scheme that underpins the \dhalo algorithm.
We use these trees only when presenting results that involve tracking the main progenitors of \eagle galaxies 
in time. We define the main progenitor as the subhalo progenitor containing the most bound particle. For the 
merger trees constructed from the hydrodynamical simulation, we ensure in post-processing that the main progenitor 
is always a dark-matter subhalo, as opposed to a fragmented clump of star and black hole particles.

\subsection{Matching haloes}

%\noalign{\smallskip}\cline{2-7}\noalign{\smallskip}

\begin{table}
%\begin{center}
  \begin{tabular}{c c c c c c}
  \hline
  $\log(M_\star \, / \mathrm{M_\odot})$ && $8-9$ & $9-10$ & $10-11$ & $11-12$ \\
  \hline
  z=0.0 & $f_{\mathrm{tot}}$ & 0.79 & 0.91 & 0.96 & 0.96  \\
        & $f_{\mathrm{c}}$   & 0.97 & 0.99 & 0.99 & 0.98  \\
        & $f_{\mathrm{s}}$   & 0.55 & 0.81 & 0.91 & 0.88  \\
  \hline
  z=2.0 & $f_{\mathrm{tot}}$ & 0.84 & 0.95 & 0.95 & 1 \\
        & $f_{\mathrm{c}}$   & 0.98 & 0.98 & 0.96 & 1  \\
        & $f_{\mathrm{s}}$   & 0.57 & 0.87 & 0.91 & 1  \\
  \hline
  z=3.85& $f_{\mathrm{tot}}$ & 0.91 & 0.97 & 0.95 & -  \\
        & $f_{\mathrm{c}}$   & 0.97 & 0.98 & 0.95 & -  \\
        & $f_{\mathrm{s}}$   & 0.66 & 0.91 & 0.85 & -   \\
  \hline
 \end{tabular}
%\end{center}
\caption{Statistics for successful matches between haloes in the dark-matter-only and reference \eagle simulation.
Binning galaxies by stellar mass in \eagle, the fractions of successful matches for central, satellite and all galaxies
($f_{\mathrm{c}}, f_{\mathrm{s}}$ and $f_{\mathrm{tot}}$ respectively)
are presented for three different redshifts. Note that at $z=3.85$, there are no galaxies with $M_\star > 10^{11} \, \mathrm{M_\odot}$.
Here, we use \eagle rather than \galform to define whether matched galaxies are considered centrals or satellites.}
\label{match_table}
\end{table}

To compare galaxies between \eagle and \galform on an object-by-object basis, we match haloes
between the reference \eagle simulation ($L100N1504$) with the corresponding dark-matter-only
simulation, following the methodology of \cite{Schaller15b}. Haloes are matched using unique dark matter
particle identifiers (IDs). For each subhalo in the reference \eagle simulation, the $50$ most bound dark 
matter particles are identified and cross-matched against the particle IDs of subhaloes in
the dark-matter-only simulation. If more than half of these particles are found to be associated with
a given subhalo in the dark-matter-only simulation, and over half of the corresponding $50$
most bound particles from that subhalo belong to the former subhalo (such that the match is
bijective), then the match is considered positive. This matching procedure is performed for a 
selection of redshifts ($z=0,2$ \& $3.9$). The matching statistics are listed in 
Table~\ref{match_table}.

\subsection{Compartmentalization}
\label{compart_section}

For each baryonic component that is included in \galform, we assign baryonic particles
to a corresponding component in \eagle. We first assign baryonic particles to a given
subhalo as described in Section~\ref{subfind_section}. The baryonic particles associated
with a given subhalo and then assigned to one of the following reservoirs:

\begin{itemize}

\item Stars-galaxy, $M_\star$ - stars associated with the galaxy.

\item Stars-ICL, $M_{\mathrm{\star, ICL}}$ - stars associated with the intra-cluster medium.

\item Halo gas, $M_{\mathrm{diffuse}}$ - diffuse circumgalactic halo gas.

\item ISM gas, $M_{\mathrm{ISM}}$ - gas in the ISM of the galaxy.

\item Ejected gas, $M_{\mathrm{ejected}}$ - gas that has been ejected (but not later reincorporated) from the subhalo.
Note that this reservoir therefore (only) includes particles that are not bound to the subhalo.

\end{itemize}

The galactic stellar component is straightforwardly defined as the stellar particles
within $30 \, \mathrm{pkpc}$ of the subhalo centre, following \cite{Schaye15}.
This aperture is used to make a distinction between stars in the
galaxy and the significant, extended intracluster light component that exists around
massive, $M_\star \sim 10^{11} \, \mathrm{M_\odot}$ galaxies in \eagle. We do not include such
an aperture for \galform galaxies at present because the corresponding massive galaxies
have much smaller half-light radii than in \eagle \cite[][]{Guo16}. When analysing
results from \eagle, we do not attempt to distinguish between stellar disc and spheroid 
components. Unless otherwise stated, all stellar properties presented from \galform
are computed by summing bulge and disc components.

In \galform, the ISM consists of two components: a rotationally supported gas disc
and a nuclear component that is associated with bursts of star formation. The
remaining gas within a given subhalo is then grouped together as a circumgalactic
halo-gas component. We define a corresponding ISM component in \eagle by selecting
gas particles within a given subhalo that are either rotationally supported against
collapse to the halo centre or are spatially coincident with the galactic stellar
component. 

We select rotationally supported gas particles as those that satisfy both 

\begin{equation}
-0.2 < \log_{\mathrm{10}}\left(\frac{2 \epsilon_{\mathrm{k,rot}}}{\epsilon_{\mathrm{grav}}}\right) < 0.2,
\label{rot_criteria}
\end{equation}

and 

\begin{equation}
\frac{e_{\mathrm{k,rot}}}{e_{\mathrm{k,rad}} + e_{\mathrm{th}}} > 2,
\label{rot_criteria2}
\end{equation}

\noindent where $\epsilon_{\mathrm{k,rot}}$ is the rotational specific kinetic energy associated
with motion orthogonal to the radial vector orientated from the gas particle to the subhalo centre\footnote{Note that there
is therefore no preferred rotation direction in our ISM selection criteria and no distinction is made
between gas particles that are corotating and those that are counter-rotating with respect to the ensemble.}. $\epsilon_{\mathrm{k,rad}}$
is the corresponding specific kinetic energy associated with radial motion. $\epsilon_{\mathrm{grav}}$ 
is the specific gravitational energy defined as $GM(r)/r$ and $e_{\mathrm{th}}$ is the specific internal
energy. 

Eqn~\ref{rot_criteria} acts to select gas particles that have the correct rotational kinetic energy to be in rotational 
equilibrium against the gravitational potential at a given radius. Eqn~\ref{rot_criteria2} acts to remove
particles with significant radial motion or with significant thermal energy. Put together, Eqns~\ref{rot_criteria}
and ~\ref{rot_criteria2} act to separate the rotationally supported ISM from a diffuse, hot gaseous halo
or from radially infalling accretion streams.

In addition to rotationally supported gas, we also select gas particles that are spatially coincident
with the stellar galactic component. Specifically, we select any dense gas 
($n_{\mathrm{H}} > 0.03 \, \mathrm{cm^{-3}}$) that is within twice the half-mass radius, 
$r_{\mathrm{1/2,\star}}$ of the stellar component of the subhalo. This acts to select dense, 
nuclear ISM gas that is typically pressure supported because of the imposed ISM equation of 
state in \eagle \cite[][]{Schaye15}.

Finally, we apply a number of radial cuts that act to remove distant, rotationally supported
material that is clearly not spatially coincident with the ISM. For inert, passive galaxies
with no centrally peaked ISM component, additional care must be taken to use radial cuts 
appropriate for these systems.

Specifically, we assign gas particles to the ISM in \eagle by applying the following selection criteria in
the following order:

\begin{enumerate}

\item We require that ISM gas must be cooler than $10^5 \, \mathrm{K}$ \textbf{or} be denser than $500$ hydrogen nuclei per cubic centimeter.
\item We require that the ISM must be rotationally supported (Eqns~\ref{rot_criteria} and \ref{rot_criteria2}) \textbf{or}
be dense ($n_{\mathrm{H}} > 0.03 \, \mathrm{cm^{-3}}$) and within $2 r_{\mathrm{1/2,\star}}$.
\item We remove ISM gas that is beyond half the halo virial radius (this step is only applied for central galaxies).
\item We remove remaining ISM gas that is beyond $2 \, r_{\mathrm{90,ISM}}$ (non-iteratively). The radius
enclosing $90 \%$ of the ISM mass, $r_{\mathrm{90,ISM}}$, is calculated after the previous selection criteria have already been applied.
\item If the galaxy has a remaining ISM gas ratio of $M_{\mathrm{ISM}} / M_\star < 0.1$, we apply a passive-galaxy correction
and remove ISM gas beyond $5 r_{\mathrm{1/2,\star}}$.

\end{enumerate}

A more detailed justification and discussion of these ISM definitions is given in Appendix~\ref{ap:ism},
along with a number of examples. Importantly, we find that the resulting ISM mass can be significantly
different compared to if the ISM is instead defined simply as star-forming gas (see also Section~\ref{threshold_section}). 
We also show that the ISM definition used here is similar at low 
redshifts to selecting mass in neutral hydrogen within a $30 \, \mathrm{pkpc}$ aperture, as used in \cite{Lagos15}.
However, at higher redshifts ($z>2$), increasingly large fractions of hydrogen in the radially 
infalling CGM are in a neutral phase. As such, the ISM definition used here starts to diverge
from taking neutral gas within an aperture.

Appendix~\ref{ap:ism} also demonstrates the important point that a simple decomposition
of baryons within haloes into distinct components does not always provide a good description of
the complex nature of what truly occurs in simulations and in reality. The separation
between the CGM and a rotation/pressure/dispersion supported ISM can be a well posed question for
many galaxies, particularly at low redshift. For some galaxies however, the distinction becomes
much less clear. Appendix~\ref{ap:ism} shows an example of a massive, high-redshift star
forming galaxy with dense, star-forming gas distributed over a significant fraction of the
halo virial radius. For such galaxies, the assumption that there is a centrally concentrated
ISM which is in dynamical equilibrium clearly starts to break down.

With gas particles belong to the ISM selected, we assign the remaining (non-ISM) gas particles associated with a given
subhalo to a diffuse halo-gas component. Finally, we also define a reservoir of gas that has 
been ejected beyond the virial radius by feedback. Note that gas that has been previously ejected 
but has since been reincorporated back inside the halo virial radius is not counted in this ejected gas reservoir.
In \galform, this reservoir is explicitly
tracked. For \eagle, we estimate the mass in this reservoir, $M_{\mathrm{ejected}}$, by taking the difference

\begin{equation}
M_{\mathrm{ejected}} = f_{\mathrm{B}} \, M_{\mathrm{H}} - M_{\mathrm{B}}
\end{equation}

\noindent where $f_{\mathrm{B}}$ is the cosmic baryon fraction and $M_{\mathrm{B}}$ is the total baryonic mass (including satellite subhaloes)
within $R_{\mathrm{200}}$. Under this approximation, gas that was prevented from ever being accreted onto
the halo in \eagle is also included in the ejected gas reservoir (assuming the halo would otherwise accrete gas at the cosmological baryon fraction).
In future work, we plan to more accurately compute the properties
of this reservoir by tracking the past/future trajectories of gas particles accreted onto haloes.

\subsection{Model calibration}
\label{calibration}

Both \galform and the \eagle simulations contain a number of model parameters that 
can be adjusted to reproduce observational constraints. The resulting calibration process 
is typically performed by hand without any statistically rigorous exploration of the model 
parameter space, although this machinery has been developed and applied for semi-analytic
models in recent years \cite[e.g.][]{Henriques13,Benson14,Lu14c,Rodrigues17}.
This calibration approach is necessary primarily 
because it is not possible at present to simulate the resolved physics of star formation and 
feedback or to model the resulting effects from first principles. 
The observational constraints on model parameters range from direct constraints like the observed 
Kennicutt-Schmidt star formation law to indirect constraints such as the luminosity function of galaxies. Broadly 
speaking, parameters relating to star formation can be calibrated directly 
\cite[][]{Schaye04, Schaye08, Lagos11a} while parameters relating to stellar feedback, SMBH accretion 
and AGN feedback are calibrated using indirect constraints. For the scientific work performed using
\galform and \eagle, the underlying philosophy regarding calibration is that a minimal set of observations
are used to adequately constrain the model parameter spaces, and that following calibration the models
can be compared to other observables with some degree of predictive power \cite[][]{Cole00, Schaye15}.

Calibration of the model parameters in the \eagle simulations is described in \cite{Schaye15} and
\cite{Crain15}. Calibration of the model parameters for the \galform model used here follows \cite{Guo16}. 
Briefly, \galform was calibrated to match the observed local luminosity functions in the $b_J$
and $K$ bands from \cite{Norberg02} and \cite{Driver12}, as well as the SMBH-bulge mass relation
from \cite{Haring04}. \eagle was calibrated to match the local stellar mass functions inferred
from observations by \cite{Li09} and \cite{Baldry12}, the local stellar mass-size relation from
\cite{Shen03} and \cite{Baldry12} and the SMBH mass versus total stellar mass relation from 
\cite{McConnell13}\footnote{\eagle was compared to the latter data set using the AGN feedback
parameter value from the \owls model. The fit was deemed satisfactory, so no calibration was necessary.
However, \cite{Booth09,Booth10} have shown that the black hole masses are determined by the subgrid AGN feedback
efficiency.}.

While \eagle and \galform were calibrated using different observational datasets, 
\cite{Trayford15} have demonstrated that \eagle agrees well with observed $u$ and $K$-band
luminosity functions. The \galform model used here predicts a similar stellar mass function
to \eagle at $z=0$, albeit with a slight deficit of galaxies around the knee. Galaxy sizes
at $z=0$ in \galform do not agree with observations (see Appendix~\ref{ap:size_mass}). Black hole
masses are comparable between \eagle and \galform at $z=0$ (see Appendix~\ref{ap:bh_sm}).

\section{Comparing stellar masses}
\label{sm_section}

\begin{figure}
\includegraphics[width=20pc]{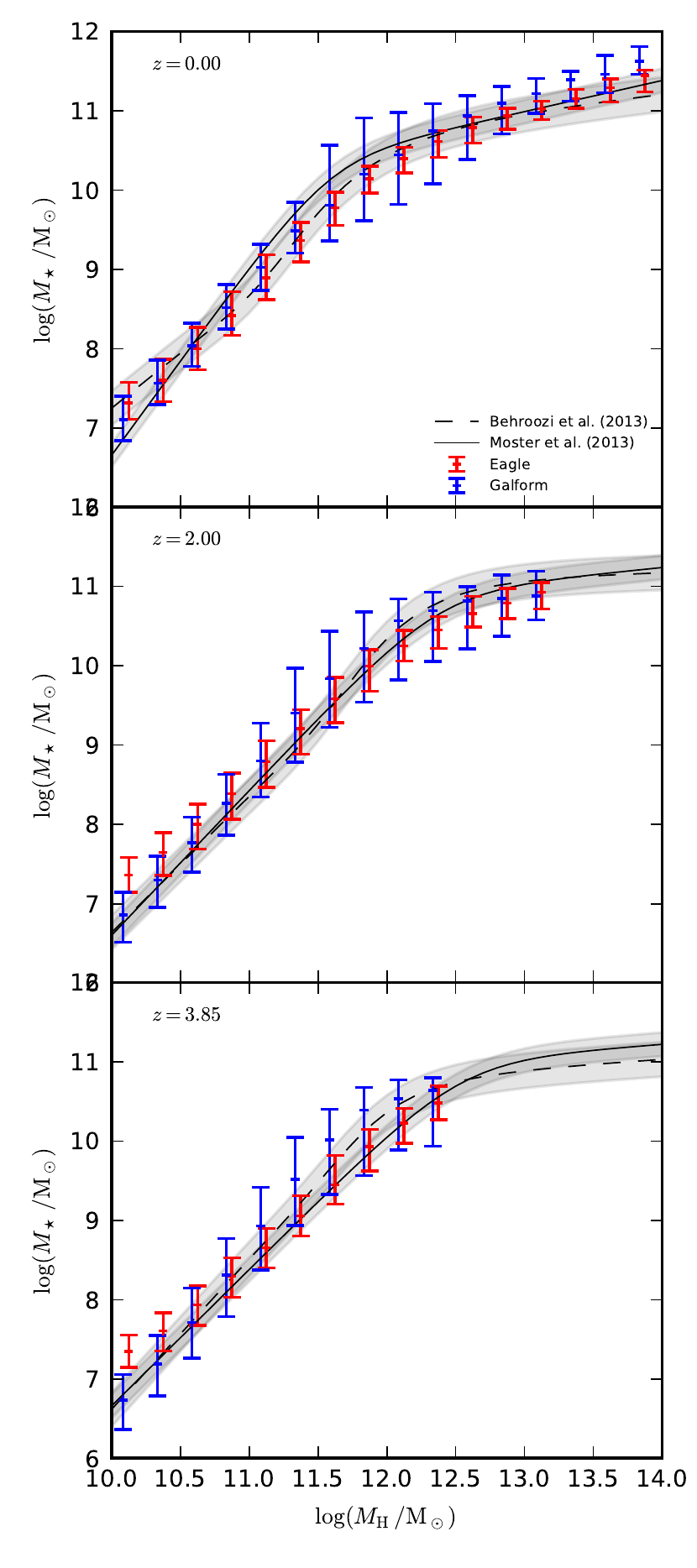}
\caption{Distribution of stellar mass as a function of halo mass for central galaxies.
Each panel corresponds to a different redshift, as labelled.
Blue points show the $16$, $50$ and $84^{\mathrm{th}}$ percentiles for the reference \galform model.
Red points show the corresponding information for the reference \eagle simulation.
%Black diagonal lines show the stellar mass that corresponds to halo mass multiplied by the cosmic baryon fraction.
Abundance matching results from \protect \cite{Behroozi13} and \protect \cite{Moster13} are shown as dashed and solid black lines.
The shaded regions show the assumed/best-fit intrinsic scatter in the distribution.}
%Halo mass is defined here as the mass enlosed within a sphere with mean density equal to $200$ times the critical density of the Universe.
%In \eagle, stellar mass is the total stellar mass within a spherical aperture of $30 \, \mathrm{kpc}$ (proper).
%In \galform, stellar mass is the total stellar disc plus bulge/spheroid, without any aperture cut.
\label{mstar_mhalo}
\end{figure}

A zeroth order comparison between the reference \galform model used here and the reference \eagle 
simulation is shown in Fig.~\ref{mstar_mhalo}, which
shows the distribution of stellar mass at a given halo mass for central galaxies. By construction,
the two models are in approximate agreement at $z=0$ where both models have been calibrated to similar
constraints. The models then diverge at $z=2,3.85$, such that \galform has a steeper relation
between stellar mass and halo mass.

Compared to \eagle, \galform has a significantly larger scatter in stellar mass at a halo mass 
$\sim 10^{12} \mathrm{M_\odot}$. For \eagle, \cite{Matthee17} showed that $0.04 \, \mathrm{dex}$ of the scatter
is connected to halo concentrations (or equivalently halo assembly times), with the remaining scatter being
uncorrelated to any of the (dark matter only) halo properties they explored.
In appendix B of \cite{Mitchell16}, we showed that the enhanced scatter at this
halo mass range in \galform is caused by the differing efficiency of SNe feedback (and hence the efficiency of
stellar mass assembly in a holistic sense) between quiescent star formation
in discs and triggered nuclear star formation associated with galaxy mergers and disc instabilities
\cite[see also the discussion in][]{Guo16}.
This differring efficiency is caused by scaling the efficiency of SNe feedback with disc circular
velocity for disc star formation and scaling with bulge circular velocity for SNe feedback associated with
triggered nuclear star formation. This in turn leads to a bimodal distribution in stellar mass at a given halo 
mass in the mass range for which the contributions to the total stellar mass from stars formed in discs 
and stars formed in nuclear bursts are comparable \footnote{The mass range for which this occurs is set by AGN feedback
in the sense that it acts to prevent late-time disc star formation from overwhelming the contribution from nuclear star
formation which tends to dominate in massive galaxies at high redshift \cite[][]{Lacey16}.}.

\begin{figure}
\includegraphics[width=20pc]{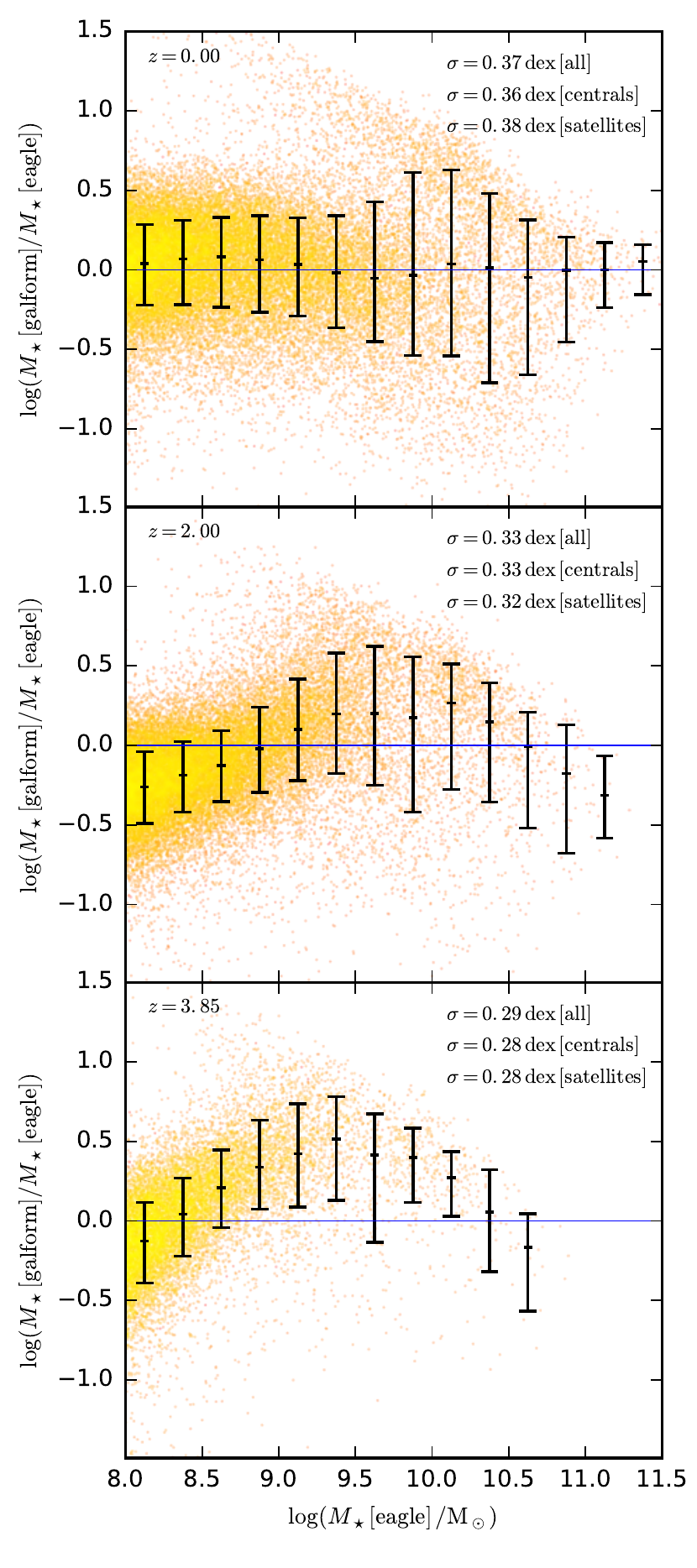}
\caption{Ratio of the stellar mass in \galform to the stellar mass in \eagle plotted as a function of stellar mass in \eagle.
Each panel corresponds to a different redshift, as labelled.
Points and error bars show the $16$, $50$ and $84^{\mathrm{th}}$ percentiles of the distribution.
$\sigma$ quantifies the $68\%$ range scatter for all, central and satellite galaxies, computed by taking the mean of the scatter for bins over the stellar mass range, $M_\star > 10^8 \, \mathrm{M_\odot}$.}
\label{stellar_mass_comp}
\end{figure}

Fig.~\ref{stellar_mass_comp} compares the stellar masses of matched individual galaxies from the two
reference models. At $z=0$, the medians of the distribution are approximately 
consistent with a unity ratio between the two models. This primarily reflects the fact that both models
are calibrated by luminosity/stellar mass function data from the local Universe. Of more interest
is the scatter in the distribution. Taking the average over stellar mass bins for $M_\star > 10^8 \, \mathrm{M_\odot}$,
the mean $1 \sigma$ scatter in the logarithmic distribution is $\sigma = 0.37 \, \mathrm{dex}$. As such,
\galform typically yields the same stellar masses as \eagle galaxies to within a factor $2.3$ at $z=0$ .
This significantly exceeds the scatter in halo mass between matched haloes from the \eagle hydrodynamical
and dark-matter-only simulations ($0.04 \, \mathrm{dex}$ at $z=0$, see Appendix~\ref{ap:halo_mass}). It is 
notable that the level of scatter is elevated for $10<\log(M_\star/\mathrm{M_\odot})<10.5$. This 
reflects the increased scatter in stellar mass at a given halo mass seen in Fig.~\ref{mstar_mhalo}
for this mass range.

At $z=2$ and $z=4$, the medians of the distribution are no longer consistent with a ratio of unity, reflecting
the different shapes of the $M_\star-M_{\mathrm{H}}$ distributions predicted by the two models
at these redshifts (see Fig.~\ref{mstar_mhalo}). The scatter around the median drops with increasing
redshift, indicating that the stellar masses of individual galaxies gradually diverge between the
models as the galaxies evolve. Decomposing the scatter between central and satellite galaxies
shows that satellites are equivalent to central galaxies in the level of agreement between the models.
We explore the underlying reasons for the overall level of scatter seen in Fig.~\ref{stellar_mass_comp} 
in Section~\ref{infall_scatter}, where we show that the implementation of gas infall and AGN feedback
in \galform leads to artificially oscillating baryonic assembly histories for individual galaxies.

\section{Star formation and the ISM}
\label{sf_section}

\subsection{Star formation threshold}
\label{threshold_section}

In \eagle, a local metallicity-dependent density threshold is used to decide which gas particles
are star-forming \cite[][]{Schaye15}. In \galform, star formation occurs in molecular gas, and
the formation of a molecular phase is explicitly computed following empirical correlations 
inferred from observations \cite[][]{Lagos11a}. Further details of the modelling are presented
in Appendix~\ref{ap:sf_th}.

\begin{figure}
\includegraphics[width=20.5pc]{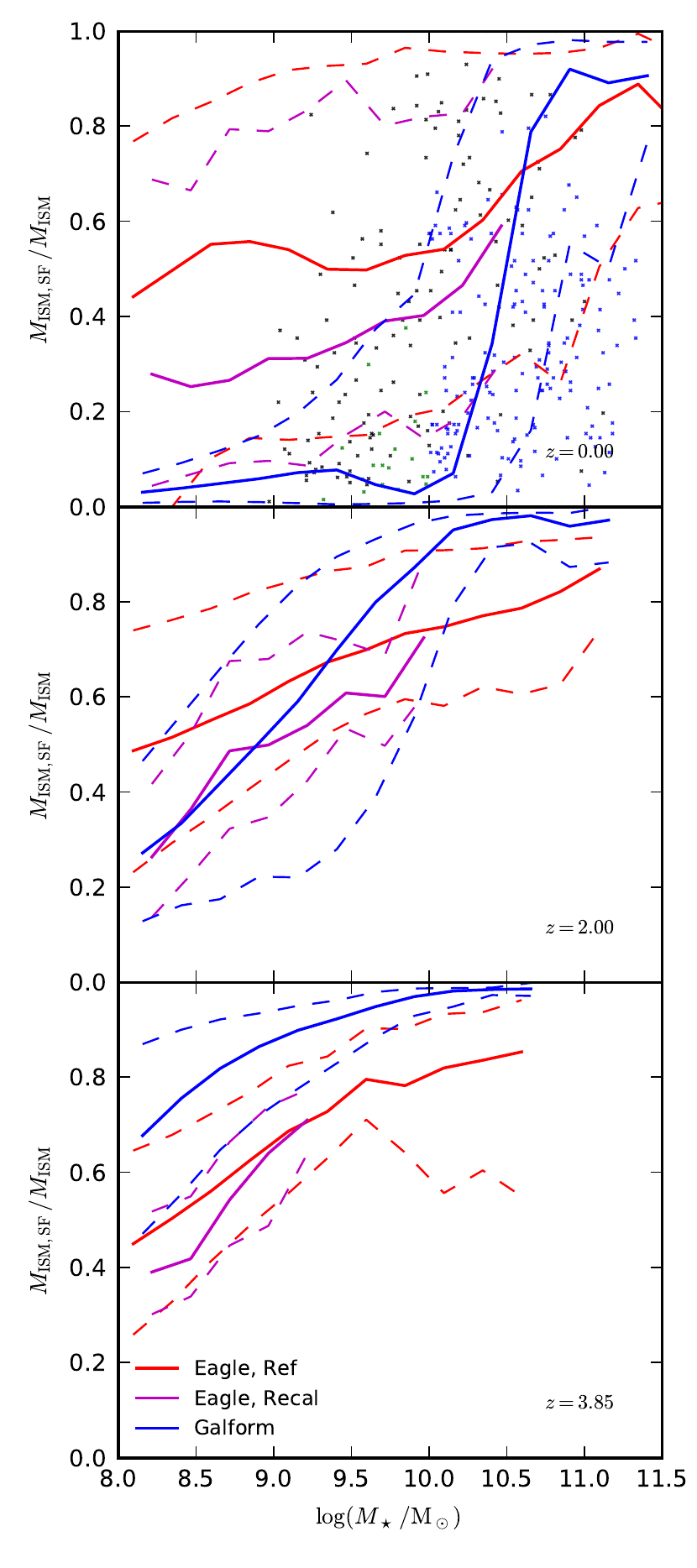}
\caption{Mass fraction of hydrogen in the ISM that is star-forming as a function of stellar mass.
Red and blue lines show the distributions from \eagle and \galform respectively.
Magenta lines show the distribution from the higher resolution, recalibrated \eagle model.
Solid lines show the medians and dashed lines show the $16$ and $84^{\mathrm{th}}$ percentiles of the distributions.
Small grey, green and blue points show respectively observational data from the HRS, ALLSMOG, GASS plus COLD GASS surveys \protect \cite[][]{Catinella10, Saintonge11, Boselli14, Bothwell14}.
Each panel corresponds to a different redshift, as labelled.}
\label{sf_threshold}
\end{figure}

Fig.~\ref{sf_threshold} shows the mass fraction of ISM gas that is actively forming
stars. In both reference models, this distribution evolves with redshift, 
reflecting the evolution of galaxy disc surface density profiles, the incidence
of disc instabilities and galaxy mergers in \galform, and the evolution in local ISM
density and metallicity in \eagle. At $z=0$, the two models display qualitatively different
behaviour. While both models predict high star-forming ISM fractions for massive
galaxies (where overall gas fractions are very low), \galform predicts significantly
lower star-forming fractions in low-mass galaxies.

In the reference \eagle simulation, the gas-phase mass-metallicity relation is
shallow at $z=0$, in tension with observational constraints which imply a 
lower gas metallicity in low-mass galaxies \cite[See figure 13 in ][]{Schaye15}.
As such, we expect that were \eagle to predict more realistic metallicities
for low-mass galaxies, the star-forming ISM fraction would be correspondingly lower.
We can test this hypothesis by considering the higher resolution recalibrated
\eagle model (magenta lines in Fig.~\ref{sf_threshold}). In this model, the 
mass-metallicity relation is steeper, in better agreement with observations 
\cite[][]{Schaye15}. Correspondingly, Fig.~\ref{sf_threshold} shows that the 
star-forming ISM fraction is smaller for this model in low-mass galaxies.

While \galform does not reproduce the observed metallicities either \cite[see][]{Guo16},
this is irrelevant for star formation because the star formation threshold has no metallicity dependence
in this model. Unlike \eagle, however, \galform predicts galaxy sizes for low-mass 
late-type (disc) galaxies that are too large compared to observations in \galform 
(the galaxy size distributions as a function of stellar mass are shown in Appendix~\ref{ap:size_mass}). 
As such, gas surface densities (for a given ISM mass)
will be unrealistically low in low-mass galaxies, potentially leading to unrealistically low molecular 
gas fractions. 
Observational data from the Herschel Reference Survey \cite[HRS,][]{Boselli14}, the APEX low-redshift legacy survey for molecular gas \cite[ALLSMOG,][]{Bothwell14}, the Galex Aricebo SDSS survey
\cite[GASS,][]{Catinella10} and the CO legacy database for GASS \cite[COLD GASS,][]{Saintonge11} indicate that is indeed the case,
with a significant number of detected galaxies with higher molecular-to-atomic gas fractions
than is predicted by \galform for stellar masses lower than $10^{10} \, \mathrm{M_\odot}$.
The data also indicates that there are massive galaxies with lower molecular-to-total ISM
gas fractions than those predicted by either \galform or \eagle (taking molecular-to-total
ratio as a proxy for the star-forming to total ISM ratios shown for \eagle).

At higher redshifts, Fig.~\ref{sf_threshold} shows that the fraction of
mass in the star-forming ISM in the two reference models comes into slightly better 
agreement. Qualitative differences remain however. At $z=2$, \galform exhibits 
a steeper trend with stellar mass. At $z=3.85$, the star-forming ISM fraction
is systematically higher by $20 \%$ in \galform at all stellar masses.

\subsection{Star formation law}
\label{sf_eff_section}

In \eagle, star-forming gas is turned into stars following a Kennicutt-Schmidt star
formation law, reformulated as a pressure law \cite[][]{Schaye08,Schaye15}.
In \galform, the star formation rate in galaxy discs is linear in the molecular
gas mass in the disc, with a constant, empirically constrained conversion efficiency \cite[][]{Lagos11a}.
Further details of this modelling are presented in Appendix~\ref{ap:sf_law}.

\begin{figure}
\includegraphics[width=20pc]{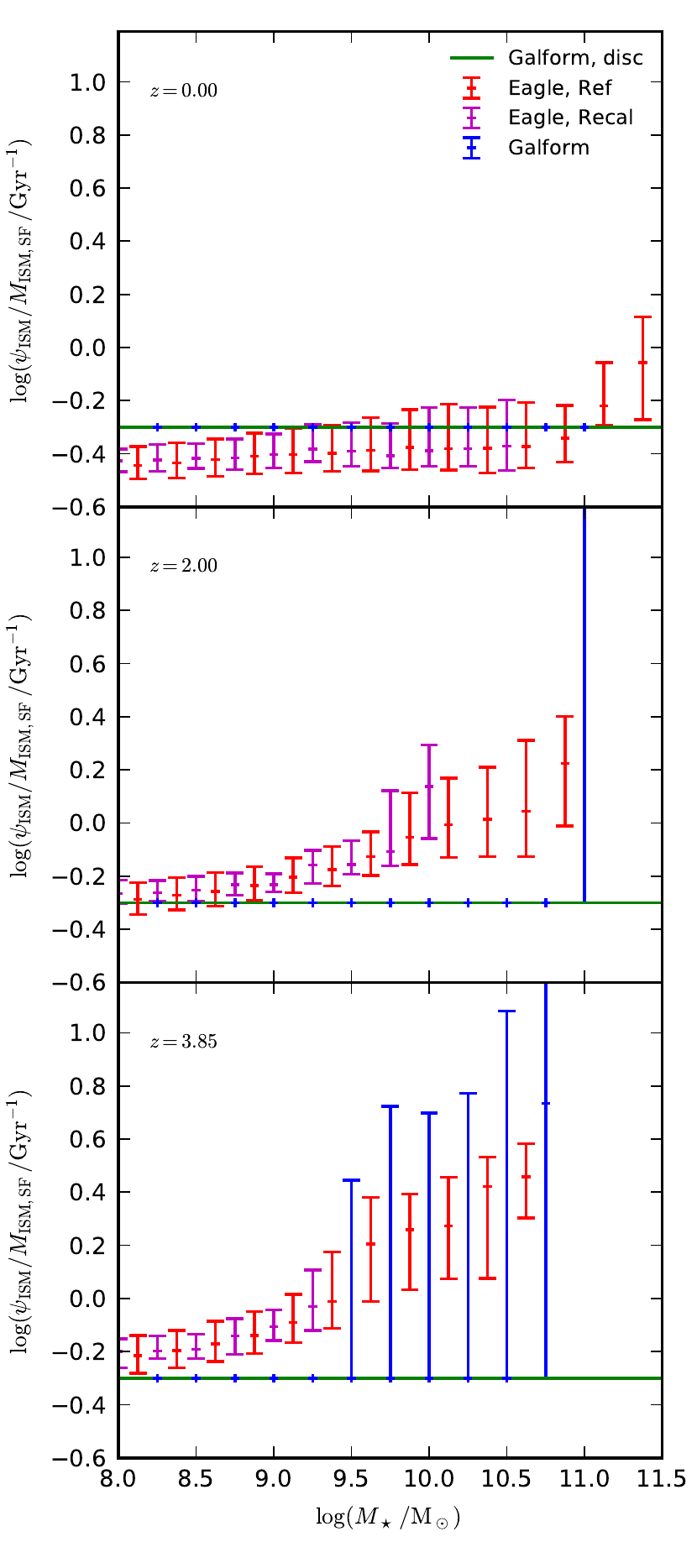}
\caption{SFR per unit star-forming ISM mass, plotted as a function of stellar mass for actively star-forming galaxies (see text for details).
Red and blue points show respectively the distributions from the reference \eagle and \galform models.
Magenta points show the distribution from the higher resolution, recalibrated \eagle simulation.
Also shown is the SFR per unit star-forming ISM mass that occurs in galaxy discs in \galform (by construction a constant, indicated by green lines),
which excludes the contribution from disc-instability or merger triggered starbursts.
Points and error bars show the $16$, $50$ and $84^{\mathrm{th}}$ percentiles of the distributions.
Each panel corresponds to a different redshift, as labelled.}
\label{sfe}
\end{figure}

Fig.~\ref{sfe} shows the SFR per unit star-forming ISM mass for actively star-forming
galaxies (which we define as specific SFR $> 0.01, 0.1, 0.1 \, \mathrm{Gyr^{-1}}$ for $z=0,2,3.85$ respectively \footnote{This
is guided by the distributions of specific SFR for \galform and \eagle shown in figure 1 of \cite{Mitchell14} and figure 7 from \cite{Guo16}}).
At $z=0$, \galform has a slightly
higher star formation efficiency but agrees with \eagle (for both the reference and recalibrated simulations) 
to within $\approx 40 \%$, except for the most massive galaxies in the simulation. At higher redshifts, the agreement
worsens as \eagle displays a significant positive trend of efficiency with stellar mass.
Star formation efficiency also increases with redshift at fixed stellar mass in \eagle
(albeit more strongly for more massive galaxies). This reflects the changing ISM conditions
for the star-forming gas in \eagle with mass/redshift. At high-redshift, the typical densities
of star-forming ISM gas increase in the simulation, accordingly increasing the gas pressure
and hence the efficiency of star formation, following Eqn~\ref{sfr_eqn_eagle} 
\cite[see figure 12 from][]{Lagos15}. Despite assuming
that star formation in galaxy discs has a fixed efficiency for star-forming gas, this effect is
somewhat accounted for in \galform by the inclusion of explicit nuclear bursts of star formation.
This elevates the net star formation efficiency of a subset of the massive galaxies at high-redshift,
albeit with a very skewed distribution compared to \eagle.

\subsection{Gas fractions}

\begin{figure}
\includegraphics[width=20pc]{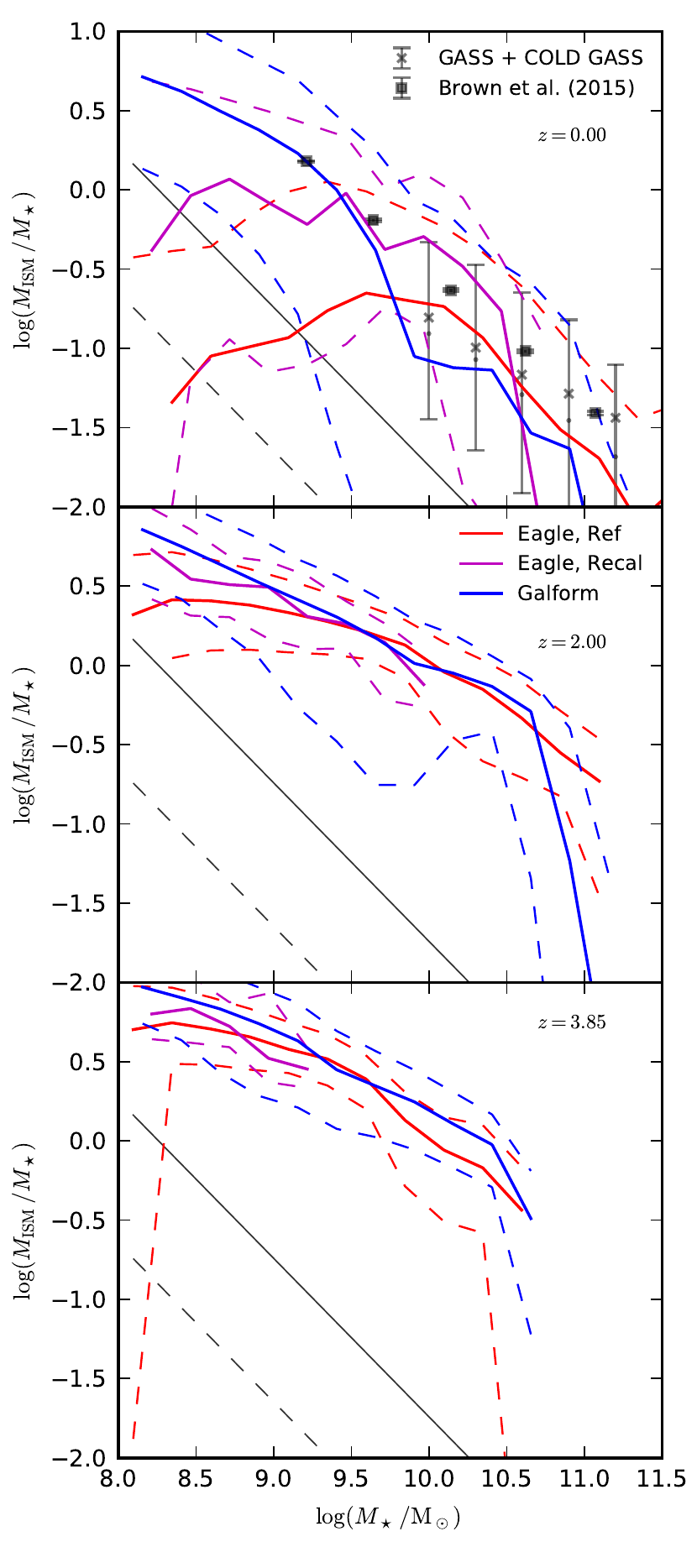}
\caption{ISM gas ratios as a function of stellar mass.
%ISM gas fraction is defined as the ratio of ISM gas mass to stellar mass in the galaxy.
Red and blue lines show the distributions from the reference \eagle and \galform models respectively.
Magenta lines show the distribution from the higher resolution, recalibrated \eagle model.
Solid lines show the medians and dashed lines show the $16$ and $84^{\mathrm{th}}$ percentiles of the distributions.
Solid grey lines mark the point below which the number of gas particles in the ISM drops below $100$ in the reference \eagle simulation.
The dashed grey lines mark the corresponding point for the higher resolution, recalibrated \eagle simulation.
Black points show gas ratios of neutral gas from the GASS and COLD GASS surveys and atomic gas ratios from \protect \cite{Brown15}.
For GASS/COLD GASS, the error bars indicate the $1$ sigma intrinsic scatter of the distribution.
Each panel corresponds to a different redshift, as labelled.}
\label{ism_fraction}
\end{figure}

The result of the differing star formation thresholds and efficiencies
(as well as the effect of accretion and outflow rates, which we do not measure 
here) are reflected in the galaxy gas-to-stellar mass ratios
($M_{\mathrm{ISM}}/M_\star$), shown in Fig.~\ref{ism_fraction}.
For galaxies with $M_\star > 10^{10} \mathrm{M_\odot}$ at $z=0$, the two 
models are consistent with each other and with the GASS and COLD GASS
surveys. At lower masses, the gas ratios
in the reference \eagle simulation drop until there are typically
only a handful of gas particles in the ISM of a galaxy of $M_\star \sim 10^8 \, \mathrm{M_\odot}$.
In stark contrast, \galform predicts that low-mass galaxies have 
much higher gas ratios, such that the gas ratio decreases 
monotonically with increasing stellar mass. The median
gas ratios in \galform are consistent with the stacked
atomic gas ratios from \cite{Brown15} (note that in this regime,
the ISM is almost entirely atomic in \galform, see Fig.~\ref{sf_threshold}).

At higher redshifts, the two models are in better agreement,
both showing the expected trend of increasing gas ratio with redshift
(although there is little difference $z=2$ and $z=4$).
Here, it appears that a higher star formation efficiency
in \eagle (Fig.~\ref{sfe}) is compensated for by a lower fraction of the ISM which
is forming stars (Fig.~\ref{sf_threshold}). We have confirmed that \eagle and \galform
are indeed very similar at these redshifts in the star formation
efficiency per unit total ISM mass (as opposed to the efficiency
per unit star-forming ISM mass shown in Fig.~\ref{sfe}).

Fig.~\ref{ism_fraction} also shows the gas ratios from the higher-resolution, 
recalibrated \eagle model. This indicates that the non-monotonic behaviour seen 
for the reference \eagle model at $z=0$ is likely a resolution effect. The 
recalibrated model shows similar non-monotonic behaviour but at a lower stellar 
mass. The recalibrated model is in better agreement with \galform and the observational
data as a result. Grey solid and dashed lines show the point below which the number
of gas particles in the ISM drops below $100$ for the reference and recalibrated
\eagle models respectively. This indicates that resolution is indeed likely to
be an issue for galaxies with $M_\star < 10^{10} \, \mathrm{M_\odot}$ 
in the reference \eagle model at $z=0$ \cite[see the discussion in ][]{Crain17}
and may also affect the higher-resolution recalibrated model for $M_\star < 10^9 \, \mathrm{M_\odot}$.

To summarise the differences between \galform and \eagle seen in this section
(Fig.~\ref{sf_threshold}, \ref{sfe} and \ref{ism_fraction}), we have demonstrated
that gas-to-stellar mass ratios are much higher in \galform than in \eagle for low-mass galaxies at $z=0$
(with an apparent connection to numerical resolution in \eagle),
but that the models are in good agreement at higher redshifts. For low-mass
galaxies at low-redshift, this difference seems to be connected to
the difference in the fraction of the ISM that is star-forming (with \galform having
much lower star-forming ISM fractions). A simple explanation for this difference 
in star-forming ISM fraction is that because low-mass galaxy sizes are significantly larger in \galform than in \eagle
at $z=0$ (see Appendix~\ref{ap:size_mass}), leading to lower gas surface densities at
at a given gas fraction and stellar mass in \galform.
At higher redshifts, the star-forming ISM fractions are less discrepant but 
\eagle has a significantly higher median star-formation efficiency than \galform for the star-forming
ISM in massive galaxies. This translates to very similar gas-to-stellar mass ratios with 
respect to \galform, partly because the differences in star-forming ISM
fraction and star formation efficiency compensate for each other, and 
possibly because the burst mode of star-formation in 
\galform does indeed compensate for the increased median efficiency in \eagle.

\section{Angular momentum}
\label{j_section}

\begin{figure}
\includegraphics[width=20pc]{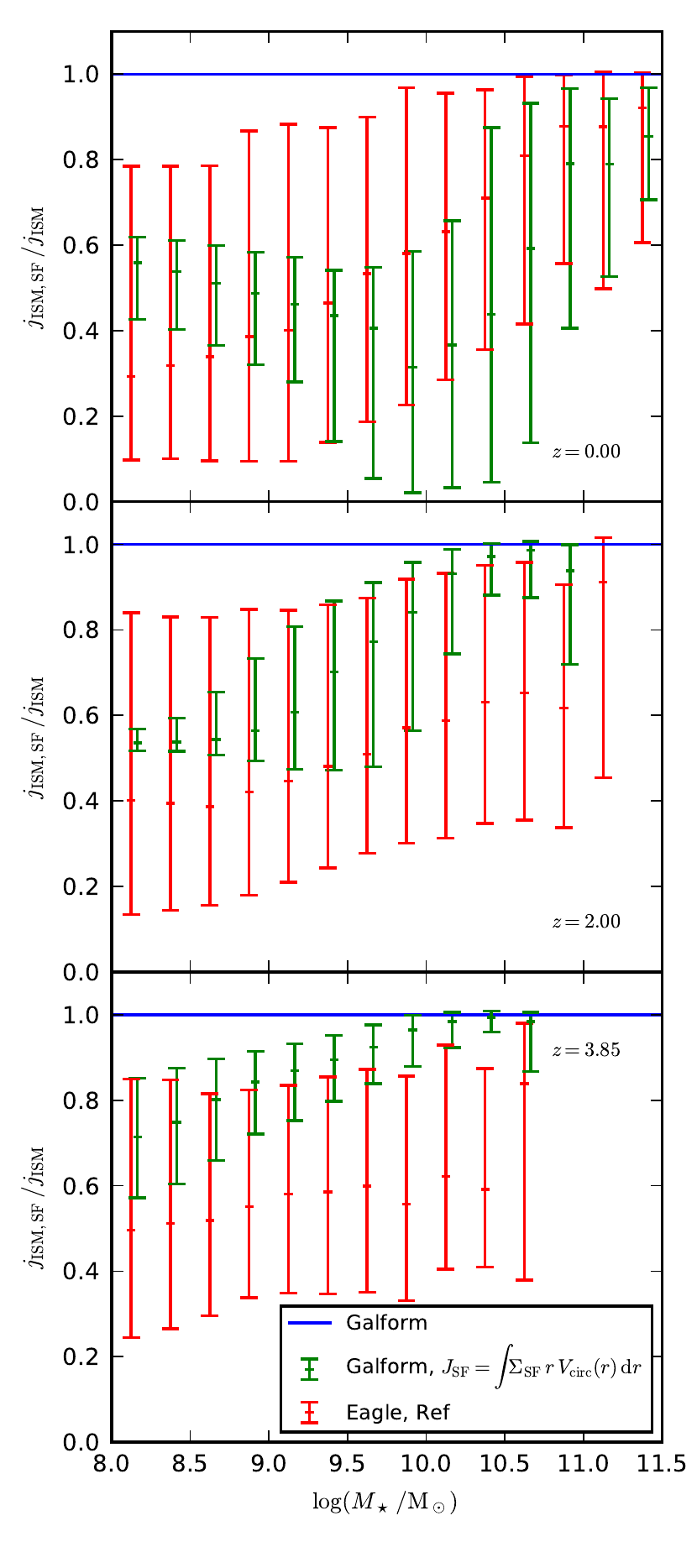}
\caption{Ratio of specific angular momenta of the star-forming ISM, $j_{\mathrm{ISM,SF}}$ to that of the total ISM, $j_{\mathrm{ISM}}$.
Red points show the distribution from \eagle.
The blue line shows the unity ratio implicitly assumed in \galform.
Green points show the distribution \galform would predict were the star-forming disc gas to have angular momentum self-consistent with the radial profile of molecular hydrogen, assuming a flat rotation curve.
Here, we make the approximation that nuclear ISM gas undergoing bursts of star formation has no angular momentum.
Points and errorbars show the $16$, $50$ and $84^{\mathrm{th}}$ percentiles of the distributions.
Each panel corresponds to a different redshift, as labelled.}
\label{jism}
\end{figure}

An important difference between \eagle and \galform concerns the angular momentum of
newly-formed stars (for full details of angular momentum modelling in \galform, 
see Appendix~\ref{ap:j}). 
In \eagle, stellar particles self-consistently inherit the angular 
momentum of the gas from which they formed. Historically, this was also
the case in older \galform models when there was no partition between atomic and molecular gas
(and hence no star formation threshold) \cite[][]{Cole00}.
Specifically, it was assumed that the gas and stars within the disc shared a common
radial scale length and correspondingly had identical specific-angular momentum.
While a simplifying assumption, this did ensure that newly-formed stars
had consistent specific angular momentum with the star-forming gas.
After the introduction of a radius-dependent partition between atomic and molecular hydrogen by \cite{Lagos11a} 
this assumption was retained. As such, newly-formed stars in our reference \galform
model have the same specific angular momentum as the total ISM gas disc, rather than 
just that of the star-forming ISM (molecular hydrogen). Given that the star-forming,
molecular ISM is more centrally concentrated than the atomic ISM under the \cite{Lagos11a}
scheme, this means that newly-formed stars have inconsistently high specific angular momentum
in \galform. 

This is explicitly demonstrated in Fig.~\ref{jism}, where we show the ratio of (magnitude of the) specific
angular momentum in the star-forming ISM to specific angular momentum in the total ISM.
Compared to the unity ratio implicitly assumed in \galform, \eagle predicts that stars form 
preferentially out of ISM with lower specific angular momentum. 
In other words, star formation is centrally concentrated in \eagle. This can be understood
as a consequence of the (metallicity-dependent) density threshold implemented in \eagle. 
(ISM gas is more likely to pass the threshold at the galaxy centre where densities
are highest). There is significant scatter in the distribution, presumably reflecting the vector
nature of angular momentum (which is ignored in \galform) being affected by complexity 
of merger events and accretion flows changing in orientation over time \cite[][]{Lagos17b,Lagos17a}.

The green points in Fig.~\ref{jism} show the ratio of specific angular momenta
that \galform would predict were it to self-consistently compute the angular momentum
content of star-forming gas from the radial profile of molecular hydrogen\footnote{We emphasise
that we have not internally modified the \galform model to perform this calculation, and as such
the distribution (and many other predictions of the model) would likely look different
if we were to do so.}. Given that it is otherwise not defined in the model, we have assumed here that 
star-forming nuclear gas present in galaxy bulges has zero (net) angular momentum, although 
in practice this choice has negligible effect on the distributions shown.
From Fig.~\ref{jism}, it is apparent that were \galform to self-consistently compute
the angular momentum of the star-forming ISM when computing the angular momentum
of newly-formed stars, then \eagle and \galform would come into better agreement on average.

\begin{figure}
\includegraphics[width=20pc]{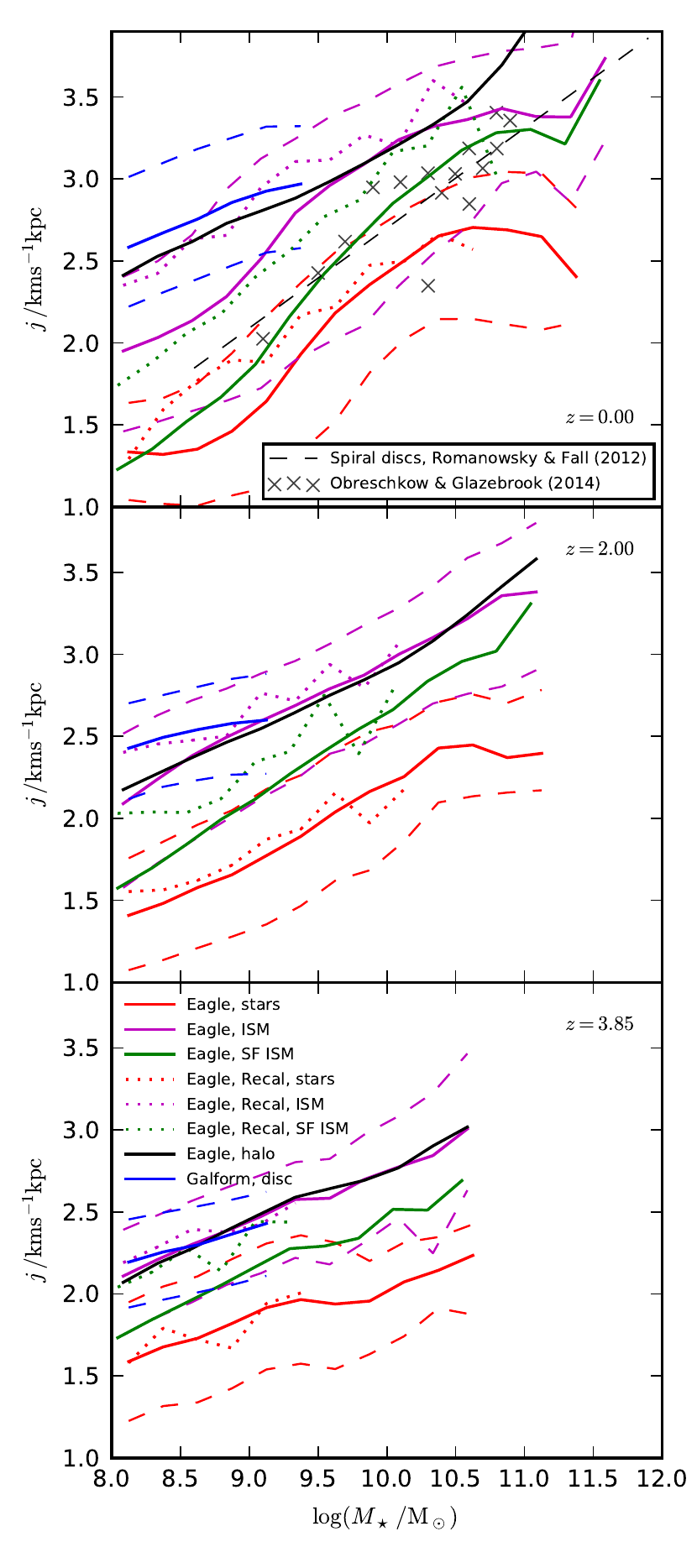}
\caption{Specific stellar angular momentum as a function of stellar mass.
Red and blue lines show respectively the distributions of stellar specific angular momentum from the reference \eagle and \galform models.
For \galform, we only show stellar mass bins that are disc dominated (such that stellar specific angular momentum is well defined in the model).
Magenta and green lines show respectively the distributions of total ISM and star-forming ISM specific angular momentum from the reference \eagle simulation.
Solid lines show the medians and dashed lines show the $16$ and $84^{\mathrm{th}}$ percentiles of the distributions.
Dotted red, magenta and green lines show respectively the corresponding median angular momenta from the higher resolution, recalibrated \eagle simulation.
Solid black lines show the median specific angular momentum of dark matter haloes at a given stellar mass in \eagle.
Grey points show galaxies from the \things survey \protect \cite[][]{Obreschkow14}.
The grey dashed line shows the power-law fit to the distribution of disc specific angular momentum from \protect \cite{Romanowsky12}.
Each panel corresponds to a different redshift, as labelled.}
\label{jstar_mstar}
\end{figure}

The impact of the assumptions regarding specific angular momentum of newly-formed stars 
is made clear in Fig.~\ref{jstar_mstar}. Focussing first on \eagle, the red 
and magenta lines show respectively the specific angular momenta of stars and the total ISM 
in \eagle. These distributions are separated by almost an order of magnitude in specific angular 
momentum at a given stellar mass. The actively star-forming ISM in \eagle (green lines) has 
intermediate specific angular momentum between the total ISM and stars. 

At intermediate stellar masses ($M_\star \sim 10^{10} \, \mathrm{M_\odot}$) at $z=0$,
the ISM in \eagle has the same specific angular momentum as the hosting dark matter haloes 
(black lines).
Interestingly, the specific angular momentum of the ISM in the \eagle reference model 
dips below the halo specific angular momentum for low-mass galaxies at $z=0$. This is
not the case at $z=2,4$ and the $z=0$ feature disappears in the recalibrated \eagle model 
(dotted magenta lines). This difference is presumably related to the much higher gas ratios 
predicted by this variant model for low-mass galaxies at $z=0$ (and hence the convergence 
issues seen in Fig.~\ref{ism_fraction}).

To make a comparison between 
\galform and \eagle in Fig.~\ref{jstar_mstar} for total specific stellar 
angular momentum (blue and red lines), we must account for the problem that \galform 
does not model the angular momenta of galaxy bulges/spheroids. This is particularly 
an issue for massive galaxies (which have high bulge-to-total ratios). For \galform, 
we therefore choose to show disc specific angular momentum rather than total specific 
angular momentum and to only show the distribution for stellar mass bins where at least 
$70 \%$ of the galaxies are disc dominated  ($M_{\mathrm{disc}} / M_{\mathrm{disc}} + M_{\mathrm{bulge}} > 0.7$). 

In contrast to \eagle, \galform (blue lines) assumes that the ISM and stars have
the same specific angular momenta in galaxy discs. Fig.~\ref{jstar_mstar} shows 
that the median stellar specific angular momentum of disc-dominated galaxies in 
\galform can be over an order of magnitude larger than in \eagle at a given stellar 
mass. 
In \galform, disc-dominated galaxies have almost the same stellar (and ISM) 
specific angular momentum as the hosting dark matter haloes (black lines). 
The corresponding galaxies in \eagle have much lower specific stellar angular momentum
than their host haloes. On the other hand, the specific angular momentum of the total ISM in \eagle
is in good agreement with the specific angular momentum of \galform discs (and so 
in agreement with the specific angular momentum of the ISM in \galform discs).

At $z=0$, we also show observational data for specific stellar angular momentum of gas-rich spiral galaxies from \cite{Romanowsky12} and
\cite{Obreschkow14}. As discussed by \cite{Lagos17a}, when selecting \eagle galaxies with high
gas fractions, \eagle agrees with the observations quite well, because gas-rich
galaxies have higher specific stellar angular momentum at a given stellar mass. % If you show a size plot, ref that result here
In contrast, \galform predicts a stellar specific angular momentum which (with some extrapolation of the observations) 
is too high for low-mass galaxies. This picture is consistent with the \galform
overprediction of low-mass galaxy sizes in the local Universe (see Appendix~\ref{ap:size_mass}).
Put together, this serves to underline that self-consistently computing the angular momentum of 
newly-formed stars from centrally-concentrated (low-angular momentum) star-forming gas in the ISM 
is likely a needed ingredient for future semi-analytic models. 

Finally, we note here that the specific angular momentum of the ISM and stars
in \eagle is (to some extent) sensitive to the assumed model parameters. For example,
increasing the normalisation of the star formation law in \eagle (see Appendix~\ref{ap:sf_law})
increases slightly the ISM
specific angular momentum (not shown here), as well as lowering the ISM-to-stars
mass fraction \cite[][]{Crain17}. A simple interpretation of this trend is that 
increasing the assumed star formation efficiency depletes more of the star-forming ISM 
(by star formation and feedback driven outflows), increasing the relative
importance of the non-star-forming ISM, which is less centrally concentrated
and so has higher specific angular momentum.

\section{Feedback}
\label{fb_section}

\subsection{Stellar feedback, AGN feedback and gas return timescales}
\label{fb_subsection}

In \eagle, stellar feedback is implemented locally by injecting thermal energy into
gas particles which neighbour young star particles \cite[][]{Schaye15}. Gas
particles are heated by a fixed temperature difference, $\Delta T = 10^{7.5} \, \mathrm{K}$,
which acts to suppress artificial radiative losses \cite[][]{DallaVecchia12}.
The average injected per supernova explosion is scaled as a function of
gas density and metallicity, ranging between $0.3$ and $3$ times the
canonical energy of $10^{51} \, \mathrm{ergs}$ (with a mean value
very close to the canonical value) \cite[][]{Crain15}. In \galform, stellar
feedback is implemented globally across a given galaxy by ejecting
gas from galaxies (and haloes) with an efficiency that scales with galaxy
circular velocity. More details of this modelling, as well as a discussion
of the energetics of stellar feedback in \galform, are given in Appendix~\ref{ap:sfb}

In \eagle, AGN feedback is implemented similarly to stellar feedback, but
with a heating temperature of $\Delta T = 10^{8.5} \, \mathrm{K}$ (an order 
of magnitude higher than for stellar feedback) and with a fixed average 
efficiency (relative to the SMBH accretion rate). In principle, this
model of AGN feedback can both heat gas in the ISM and in the halo,
and remove gas entirely from haloes. In \galform, AGN feedback acts
only to prevent gas infall from the halo onto galaxy discs, and does
not eject gas from haloes. AGN feedback is activated in \galform
if the SMBH injects sufficient energy to balance radiative cooling
in the halo, and if the halo is considered to be in a quasi-hydrostatic
state. Further details of this modelling, and of the modelling
of SMBH seeding and growth, are presented in Appendix~\ref{ap:agn}.
We compare the \galform and \eagle distributions of black hole mass
as a function of stellar mass in Appendix~\ref{ap:bh_sm}, where
we show that \galform does not predict the steep dependence on
stellar mass predicted by \eagle at $z=2,4$.

\begin{figure}
\includegraphics[width=20pc]{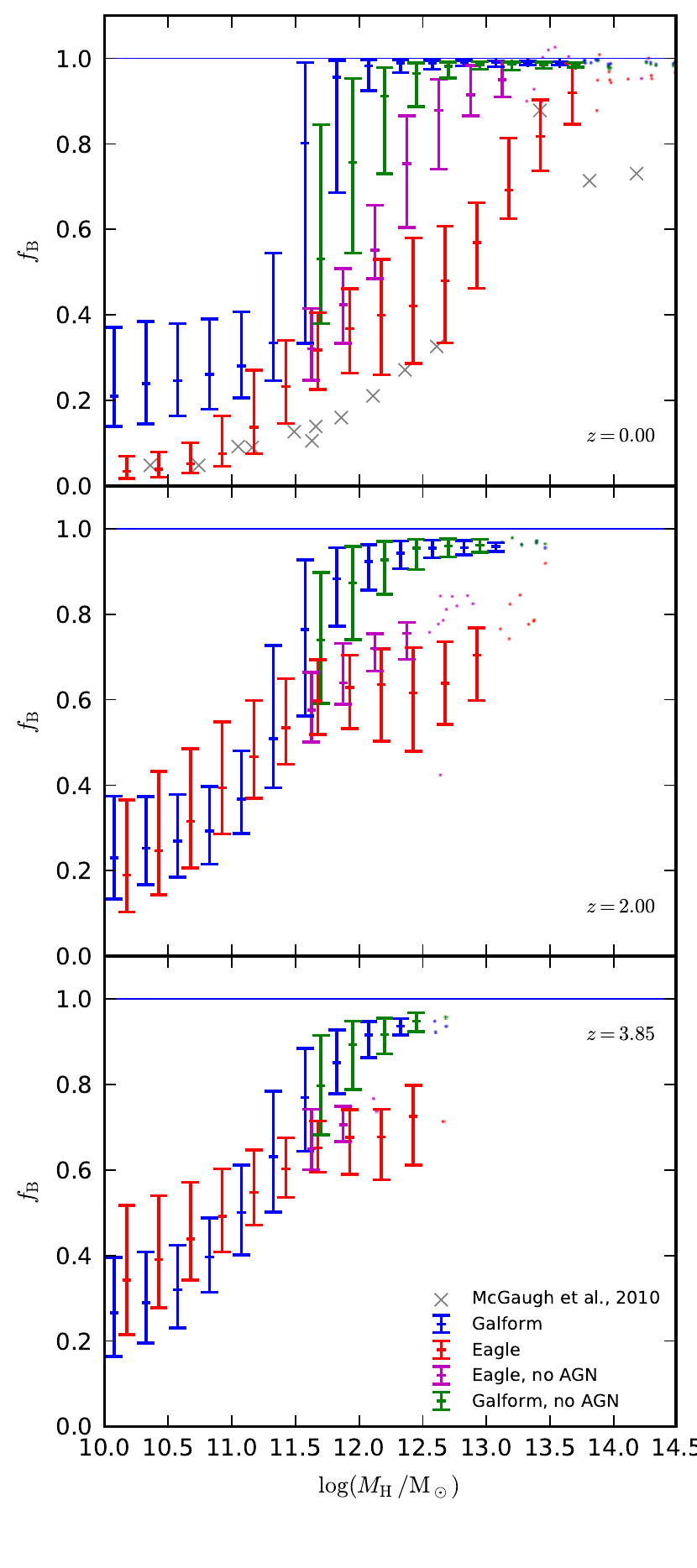}
\caption{Baryon fractions within the halo virial radius for central galaxies as a function of halo mass.
Baryon fractions are defined relative to the universal baryon fraction, with a value of unity (blue horizontal line) indicating that the halo contains the universal baryon fraction.
Red and blue points with error bars show the distributions from the reference \eagle and \galform models respectively.
In the mass range where they start to differ from the reference models, green and magenta points with error bars show 
the distributions from \galform and \eagle respectively for models that do not include AGN feedback.
The points with errorbars show the $16$, $50$ and $84^{\mathrm{th}}$ percentiles of the distributions.
The small coloured points without error bars show individual central galaxies in bins containing few galaxies.
Grey crosses show the baryonic inventory compilation from \protect \cite{McGaugh10}. 
Each panel corresponds to a different redshift, as labelled.}
\label{fbaryon}
\end{figure}

Arguably, the most important uncertainty in semi-analytic galaxy formation models
is the fate of outflowing gas that is ejected from galaxies by feedback 
\cite[e.g.][]{Henriques13,Mitchell14,White15,Hirschmann16}. In \galform, this
ejected gas is placed into a distinct reservoir and is assumed to return 
to the diffuse gas halo over a halo dynamical time (see Appendix~\ref{ap:ret}). While the spatial
location of this ejected gas is not formally defined, we choose to the
interpretation that this gas is outside the halo virial radius (see Appendix~\ref{ap:ret}).
Given that the halo dynamical time is always approximately equal to $10 \%$ of the age
of the Universe at a given epoch (independent of halo mass), ejected gas is rapidly
reincorporated back into the diffuse gas halo in \galform. This rapid gas cycling
(infall timescales from the diffuse halo onto the disc are also typically of order
a halo dynamical time) forces the model to employ very large mass-loading
factors in order to explain the low observed efficiency of cosmic star formation.
This is common to some other semi-analytic models \cite[e.g.][]{Springel01}, although 
see also \cite{Somerville08,Hirschmann16}.

Importantly, it is assumed in \galform that AGN feedback does not eject gas from
galaxies (or from haloes), and instead acts only to suppress radiative cooling from
the diffuse gas halo. Given the fairly short reincorporation timescale assumed in \galform,
this means that the baryon cycle (infall, ejection, reincorporation) effectively ceases and the halo baryon fraction rapidly 
reaches the cosmic mean once AGN feedback becomes active in a given halo \cite[although see][for alternative schemes where AGN can eject gas from haloes]{Monaco07,Bower08,Somerville08,Bower12}.

In hydrodynamical simulations, gas flows are predicted locally by self-consistently
following gravity and hydrodynamics. As such, they do not make any
explicit assumptions for gas return timescales. Implicitly however,
simulations set the return time through the details
of the subgrid models for feedback. In the case of \eagle, 
outflowing particle trajectories will be sensitive to the assumed feedback
heating temperature.
In hydrodynamical simulations without an explicitly assumed wind 
speed (and in reality), gas is presumably ejected from galaxies with a broad distribution of energies \cite[e.g.][]{Christensen16}. 
Correspondingly, it is to be expected there will be a distribution of return timescales
among gas particles associated with a wind, ranging from short timescales (less than a Hubble
time, as in \galform) to timescales much longer than a Hubble time, such that the
gas effectively never returns \cite[][]{Oppenheimer10,Christensen16,AnglesAlcazar16,Crain17}. 
It is not clear therefore whether the total return rate should be linear in the
mass of the ejected gas reservoir (as assumed in \galform, Eqn~\ref{ret_time}),
or whether the return rate will tightly correlate with halo properties, as is
assumed in semi-analytic models.

\subsection{Halo baryon fractions}
\label{fb_section}

While we do not measure mass-loading factors or gas return timescales from \eagle for this
study (we defer this to future work), an indirect measure of the efficiency
of these processes is given simply by the mass fraction of baryons within 
the virial radii of dark matter haloes. Fig.~\ref{fbaryon} shows the 
baryon fractions as a function of halo mass for a range of redshifts.
For all models, the baryon fractions rise from low values in small
haloes to high values in massive haloes, implying that to first
order the baryon cycles in the two models are similar. However, in detail 
there are various qualitative differences. 

At a characteristic halo mass of $M_{\mathrm{H}} \approx 10^{12} \, \mathrm{M_\odot}$, 
baryon fractions in \galform rapidly approach the cosmic mean. This 
transition is significantly more gradual in the reference \eagle
model, such that the baryon fraction approaches the cosmic mean
only in galaxy clusters ($M_{\mathrm{H}} \approx 10^{14} \, \mathrm{M_\odot}$).
For the models without AGN feedback (green and magenta points), 
there is still a difference between the two models for $M_{\mathrm{H}} \geq 10^{12} \, \mathrm{M_\odot}$, indicating
that SNe feedback is more effective in \eagle than in \galform
in massive haloes (either because more gas is ejected or because
gas takes longer to return). When AGN feedback is included,
\eagle and \galform show divergent behaviour. AGN feedback acts to reduce
the baryon fractions (either by direct mass ejection or by preventing
primordial accretion or gas return) in massive haloes in \eagle.
In \galform, AGN feedback does not eject gas and instead suppresses
gas cooling, which in turn acts to suppress future SNe-driven outflows,
resulting in higher baryon fractions than the no-AGN case.

In practice, suppressing cooling and ejecting gas from haloes will
both act to suppress star formation. Accordingly, the differing
baryon fractions in massive haloes between \eagle and \galform will 
not necessarily result in differing stellar mass assembly histories. 
The ejection/non-ejection of baryons by AGN feedback is relevant to the predicted 
X-ray properties of massive haloes however, as explored in \cite{Bower08}.

In lower-mass haloes, below the regime where AGN feedback plays a role,
the two models come into better agreement but still show a different 
evolution with redshift. At $z=4$,
SNe appear to be slightly more efficient in removing baryons in
\galform than in \eagle for the lowest-mass haloes shown. At $z=2$, the models are 
very comparable and by $z=0$, \eagle has a lower baryon fraction at 
a given halo mass in low-mass haloes.

At $z=0$, Fig.~\ref{fbaryon} also shows the baryon inventory
compilation from \cite{McGaugh10}, corrected to be consistent with
halo masses defined relative to $200$ times the critical density of
the Universe. Both \galform and \eagle agree with the basic qualitative
trend of increasing baryon fraction with halo mass \cite[see][to see that this is not trivially the case in simulations]{Haider16}.
For low-mass haloes ($M_{\mathrm{H}} < 10^{13} \, \mathrm{M_\odot}$), the baryonic mass estimates
presented in \cite{McGaugh10} include contributions only from
stars and the ISM, neglecting any CGM contribution (and so should
be considered lower limits). In both \galform and \eagle, a significant 
part of the baryonic mass always belongs to the CGM component, even in lower mass halos
(see Section~\ref{mass_dep_section}).

For the more massive haloes shown in the \cite{McGaugh10} compilation, 
baryonic masses are instead inferred from X-ray measurements of galaxy 
groups and clusters. Here, a detailed self-consistent comparison with 
X-ray observations that takes into account observational biases and 
systematics has not been performed (and is beyond
the scope of this paper). However, more detailed comparisons of \eagle and \galform to group and cluster
X-ray measurements have been presented in \cite{Bower08}, \cite{Schaye15} 
and \cite{Schaller15a}. For \galform, \cite{Bower08} found that the
fiducial \galform model (similar to the one shown here) significantly overpredicts
the X-ray emission from galaxy groups, and that this could be resolved
in a variant model by expelling hot gas from haloes with AGN feedback. This simple picture
is consistent with how \eagle behaves. \cite{Schaye15} and \cite{Schaller15a}
have compared \eagle to the X-ray content of galaxy groups and clusters, using
the methodology of \cite{Brun14} to perform a self-consistent comparison.
They find that the reference \eagle model shown here overpredicts the hot
gas content of galaxy groups by about $0.2 \, \mathrm{dex}$, and that this can be resolved by increasing
the heating temperature for AGN feedback. The simple comparison shown
here in Fig.~\ref{fbaryon} is consistent with this picture.

To summarise, we have shown in this section that the different implementations of SNe and AGN feedback
in \galform and \eagle lead to similar baryon fractions in lower-mass haloes,
but very different baryon fractions in group-scale haloes. This is not necessarily important
for stellar mass assembly but will have a strong impact on the X-ray properties
of galaxy groups \cite[][]{Bower08}.

\section{Galaxy evolution and the baryon cycle}
\label{cycle_section}

\begin{figure*}
\begin{center}
\includegraphics[width=40pc]{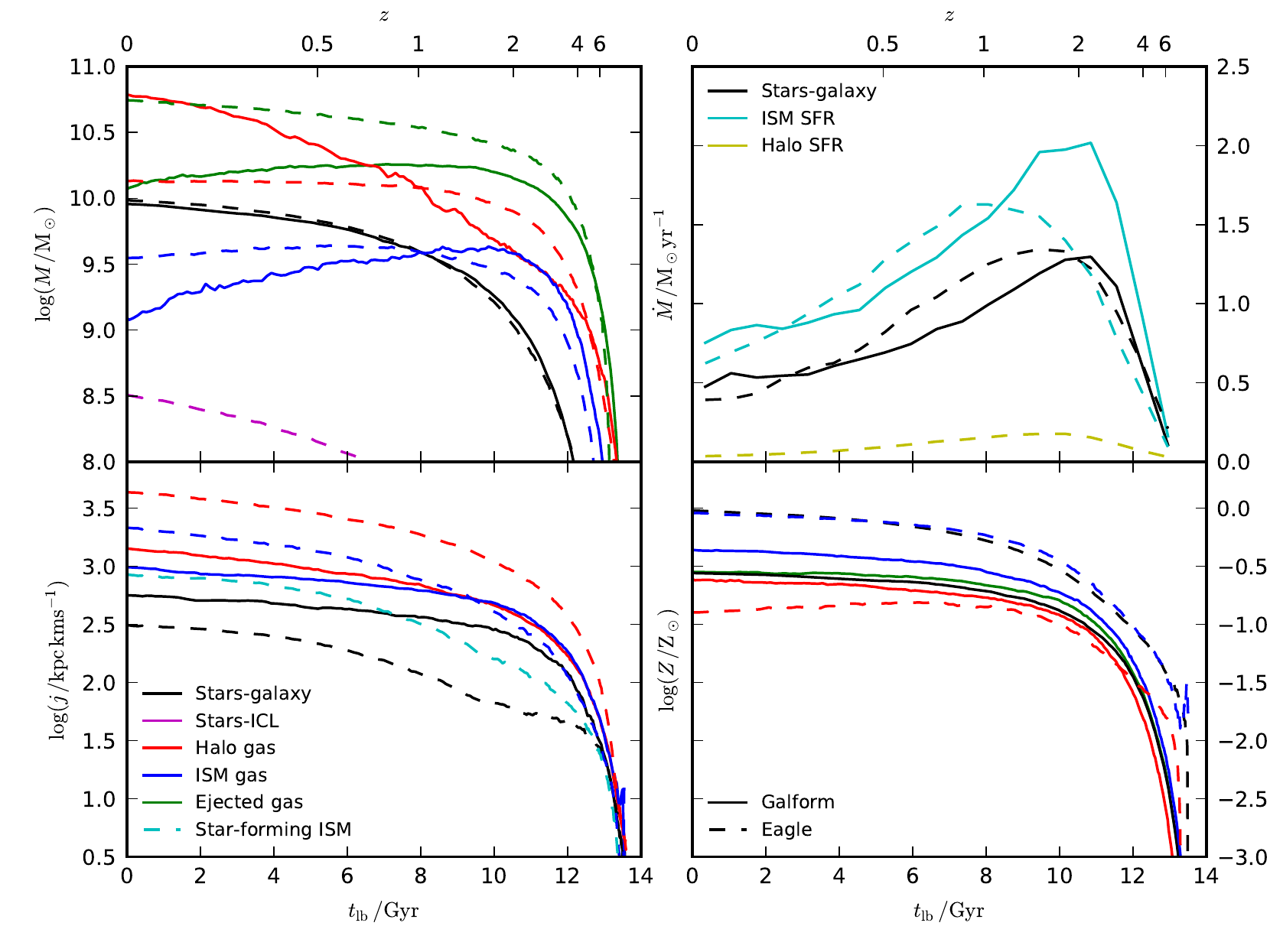}
\caption{Baryonic assembly histories as a function of lookback time for central galaxies with stellar masses in the range, $9.75<\log(M_\star/\mathrm{M_\odot})<10.25$, at $z=0$.
Here, the main progenitors of galaxies selected at $z=0$ are traced backwards in time.
Different line colours correspond to different baryonic components, as labelled.
Solid lines show the average histories from \galform.
Dashed lines show the corresponding histories from \eagle.
{\it Top-left:} Median mass in different baryonic components.
{\it Top-right:} Mean time derivative of the stellar mass (black), and mean ISM (cyan) and non-ISM (halo SFR, yellow) star formation rates.
Note that this panel is shown on a linear scale.
{\it Bottom-left:} Median specific angular momentum in different baryonic components.
For \galform, we set the (otherwise undefined) specific angular momentum of galaxy bulges to zero.
The average stellar disc specific angular momentum in \galform is identical to that of the ISM (solid blue line).
For \eagle only, we also show the specific angular momentum of the star-forming ISM in this panel (cyan dashed line). 
{\it Bottom-right:} Median metallicity in different baryonic components.
}
\label{tree_plot}
\end{center}
\end{figure*}

Here, we explore how the physical processes discussed in previous sections
(star formation, feedback and gas cycling) shape galaxy evolution by
presenting the baryonic assembly histories of galaxies in \eagle and \galform.

The time evolution of the ejected gas reservoir discussed in Section~\ref{fb_subsection}
is shown explicitly in 
Fig.~\ref{tree_plot}, in this case tracing the main progenitors
of central galaxies with $9.75<\log(M_\star/\mathrm{M_\odot})<10.25$ at $z=0$.
Green lines show the median ejected gas mass for the two models. 
Matched at this stellar mass at $z=0$, central galaxies in
\eagle (dashed lines) on average have a higher mass fraction in the ejected reservoir
at all redshifts than in \galform, and the mass in the ejected
gas reservoir increases monotonically with time. 
In contrast, \galform (solid lines) predicts that the 
median mass in the ejected reservoir rises with time up until $z \approx 1.5$, before
steadily declining until $z=0$ (note the ejected reservoir does not include ejected gas
which has been subsequently reincorporated). This is a result of the short return
timescale assumed in \galform, whereby the ejected gas reservoir closely
traces the evolution of mass in the ISM (solid blue line). In \eagle,
the evolution of the ISM mass (dashed blue line) is fairly steady
with time, rising to a peak and then modestly declining until $z=0$.
The contrast between the evolution of the ISM (slow decline) and
ejected gas reservoir (gradual rise) for $z<1$ indicates that the
net return timescale is likely longer in \eagle than is assumed
in \galform. This interpretation is supported by \cite{Crain17}, who
saw no evidence for gas particles directly heated by feedback returning to galaxies
at later times in \eagle (after inspecting past phase diagrams of gas particles
selected at a given epoch).

The stellar mass assembly histories (black lines) are in very good
agreement between the two models in this mass bin. Such a level of agreement
does not extend to any of the gas reservoirs however. This serves
to underline that the stellar assembly histories of galaxies
are not enough to constrain the feedback processes in galaxy
formation models. In particular, the median mass in the diffuse 
gas halo is a factor $4$ larger on average in \galform compared 
to \eagle. Importantly, the full distribution of masses in the 
diffuse gas halo (not shown) is significantly wider in \galform. 
We return to this issue in Section~\ref{infall_section} when discussing the 
infall/cooling model used in \galform. 

The upper-right panel of Fig.~\ref{tree_plot} shows the star formation
(cyan) and stellar mass assembly histories (time derivative
of the stellar mass, shown in black) of the same selection
of galaxies. The peak of star formation is slightly
later in \eagle with respect to \galform, which leads to better
agreement with the observed decline in specific-star formation
rates \cite[][]{Mitchell14,Furlong15}. In \cite{Mitchell14},
it was demonstrated that such a delay in the star formation
peak is only possible by introducing a very strong redshift
evolution in the gas-return timescale or mass-loading factor,
significantly beyond what is possible in the standard
\galform parameter space.
Interestingly, there is also a small contribution in \eagle 
from star formation that takes place outside the ISM (yellow 
line), a possibility that is not considered in \galform. 
Visually (see Fig.~\ref{z2_monster}), we find that radially-infalling star-forming gas 
can start to fragment and form stars before settling into 
rotational equilibrium closer to the halo centre.

The lower-left panel of Fig.~\ref{tree_plot} shows the evolution
in specific angular momentum of different baryonic components.
Specific angular momentum increases monotonically with cosmic
time for all components in both models, in accordance with tidal
torque theory \cite{Catelan96}, \citep[see][for an analysis of 
in the context of \eagle]{Lagos17a}.
Interestingly, there is a much
greater level of segregation in specific angular momentum
between different baryonic components in \eagle compared
to \galform. In \galform, angular momentum is conserved
for gas infalling onto galaxies. Furthermore, it is assumed
that outflowing gas ejected by SNe feedback has the same 
specific angular momentum as the overall ISM. Consequently,
the ISM (solid-blue line) has very similar specific angular 
momentum to the diffuse gas halo (solid-red line) as it evolves. 
The stellar specific angular momentum (solid black
line) is not fully defined in \galform (see Section~\ref{j_section})
because the angular momentum of galaxy bulges is not tracked. We choose to set 
the bulge angular momentum to zero. The resulting stellar specific angular momentum 
should therefore be regarded as a lower limit (as bulges/spheroids do rotate).
Even as a lower limit however, the stellar specific angular momentum in \galform
is still significantly higher than in \eagle.

In \eagle, the ISM and the stellar components abruptly decouple 
in specific angular momentum at redshift $\approx 6$, after
which the ISM has a factor $\approx 6$ larger specific angular momentum.
The transition redshift marks the point at which the
star-forming ISM (dashed cyan line) starts to decouple from the 
total ISM. It also marks the point at which the previously
formed stellar mass is significant enough that the past 
average specific angular momentum (represented by the stars)
drops below the ISM value. 

Another interesting feature of the lower-left panel of Fig.~\ref{tree_plot}
is that the halo gas specific angular momentum (dashed red line) in \eagle 
is positively offset with respect to the ISM (by a factor $\sim 2.5$ at $z=0$).
This behaviour is not predicted to the same extent by \galform, and not at all
for $z>1$. This closer coevolution of the ISM and halo gas in \galform
stems first from the assumption that angular momentum of gas is conserved as 
it condenses from the halo component onto a galaxy disc. Secondly, infall
rates onto galaxy discs in \galform are high enough in the non-quasi-hydrostatic 
regime that all of the halo gas at a given epoch will be accreted onto the
disc. The segregation of specific angular momentum between these components in 
\eagle is therefore suggestive that angular momentum is either not exactly 
conserved for infalling material \cite[see][for a similar conclusion]{Stevens17}, 
or that infalling material is preferentially low-angular momentum compared to the overall 
halo gas reservoir. It also seems likely that infall rates may be lower
than in \galform, such that more radially distant, high angular momentum 
gas is never accreted onto galaxies. 

The lower-right panel of Fig.~\ref{tree_plot} shows the
evolution in metallicity for different baryonic components.
\galform predicts that the median metallicity in each
component closely traces each of the others with the median
ISM metallicity positively offset by $60 \%$ from stellar
metallicity at $z=0$. In contrast, \eagle predicts that
the ISM and stellar components have almost identical metal
content at $z=0$ and co-evolve very closely (the upturn at
high-redshift in the ISM metallicity is caused by galaxies
dropping out of the sample as they can no longer be identified
by the halo finder, with the remaining galaxies probably being
unusually metal enriched at these redshifts). \eagle also
predicts that the diffuse halo gas component is negatively
offset in metallicity by a factor that grows fractionally with
time, reaching a factor $\approx 7$ by $z=0$.

In \galform, metals are exchanged between different
components such that they linearly trace the total baryonic 
mass exchange. As such, it is assumed that the metal-loading
factor (ratio of metal ejection rate from the ISM to the
rate with that ISM metals are locked into stars by star formation) is the
same as the mass-loading factor. This need not be the case
 \cite[e.g.][]{Creasey15, Lagos13}. Furthermore, \galform assumes
that newly-formed stars form with the metallicity of the total
ISM component, neglecting any possible radial gradients that could
lead to differences between the average
metallicity of star-forming gas and total ISM. Combined
with the difference in mass exchange rates implied by the
contrasting predictions shown in the upper-left panel of 
Fig.~\ref{tree_plot}, it is therefore somewhat challenging
to interpret the differences in metal evolution between
the two models. That the diffuse gas halo metallicity is
significantly lower in \eagle than the ISM and stellar components
does suggest that considerably less baryon cycling is taking place 
compared to \galform, suggesting a longer gas return timescale.

\subsection{Mass dependence}
\label{mass_dep_section}

\begin{figure*}
\begin{center}
\includegraphics[width=40pc]{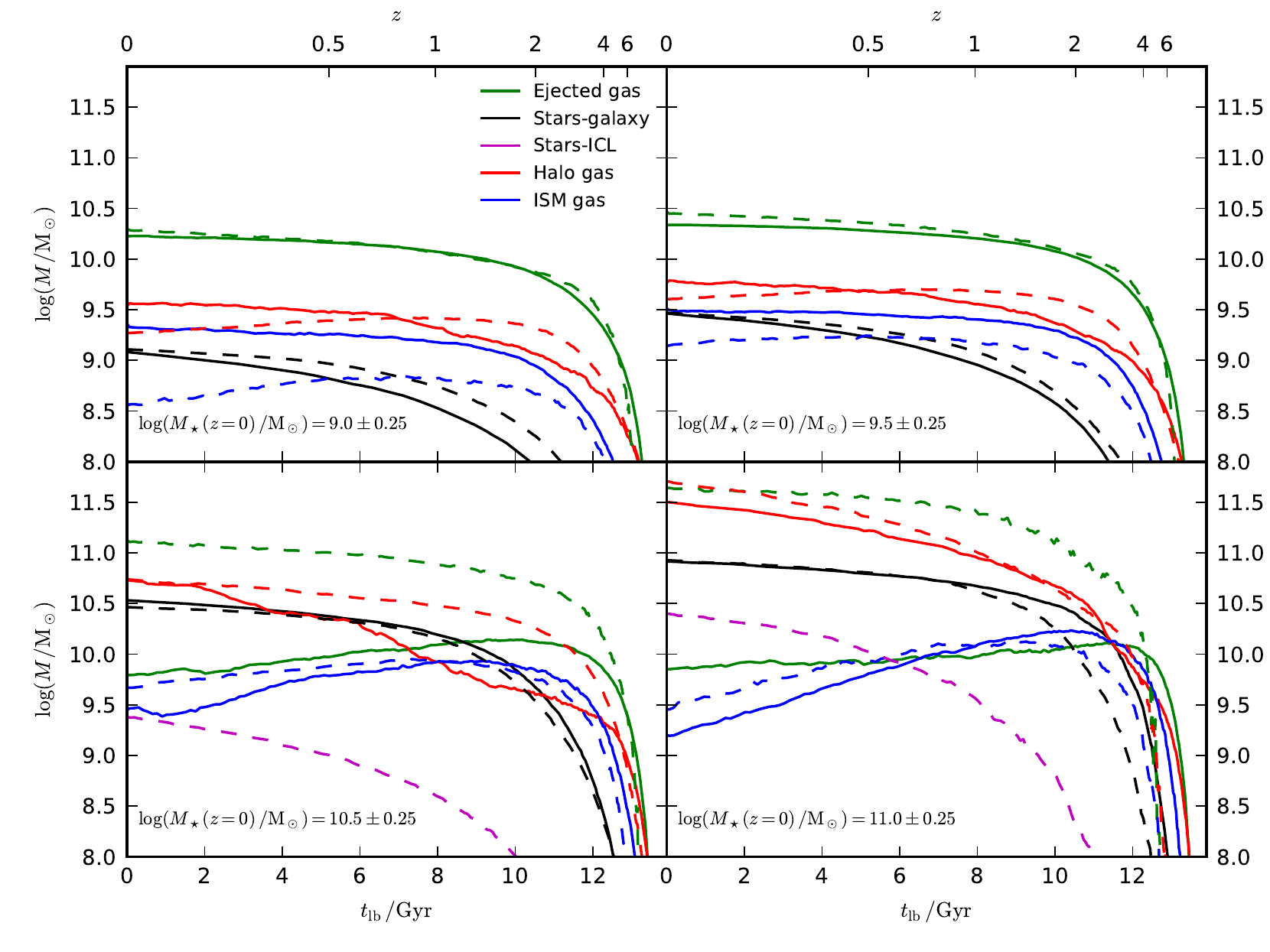}
\caption{Median baryonic mass assembly histories for central galaxies.
Each panel corresponds to a different stellar mass bin selected at $z=0$, as labelled.
Line formatting follows Fig.~\ref{tree_plot}.}
\label{tree_plot_mbins}
\end{center}
\end{figure*}

Fig.~\ref{tree_plot_mbins} shows the same information as the upper-left 
panel of Fig.~\ref{tree_plot} but plotted for a range of stellar mass bins. 
Overall, the two models come into better agreement for lower mass galaxies 
($M_\star = 10^9, 10^{9.5} \, \mathrm{M_\odot}$) but increasingly disagree
for more massive galaxies ($M_\star = 10^{10.5}, 10^{11} \, \mathrm{M_\odot}$).
For the lower mass galaxies, the ejected gas reservoir dominates the mass budget
in both models. The stellar mass in these galaxies forms later in \galform and
\galform contains systematically higher ISM content at all redshifts, presumably
resulting in more prolonged star formation histories.

For more massive galaxies (lower panels), the two models quickly start
to diverge at high-redshift in the mass of the ejected gas reservoir. 
In \galform, the ejected gas reservoir mass peaks at $z \approx 3$ before steadily declining
down to $z=0$. In \eagle, the ejected gas reservoir mass does not peak and continues to rise until
low-redshift and is always comparable to, or greater in mass than the diffuse halo gas
component. In \galform, the diffuse halo component completely dominates over the ejected
gas reservoir by $z=0$. As shown in Fig.~\ref{fbaryon}, this primarily reflects the different
implementations of AGN feedback in the two models.

It is also notable that the total baryonic mass
is significantly larger in \eagle than in \galform for the higher mass stellar mass bins shown. 
In this stellar mass range, the
$M_\star$ - $M_{\mathrm{H}}$ relation is shallow, such that a small difference
in stellar mass leads to a large difference in halo mass (see Fig.~\ref{mstar_mhalo}).
Given that we match galaxy samples here at a fixed stellar mass, the difference in total
baryonic mass for the two models is therefore primarily driven by the difference
in host halo mass at a fixed stellar mass. The difference in halo mass definition also 
contributes (see Appendix~\ref{ap:halo_mass}).

\subsection{The mass-loading / return-time degeneracy}

\begin{figure}
\includegraphics[width=20pc]{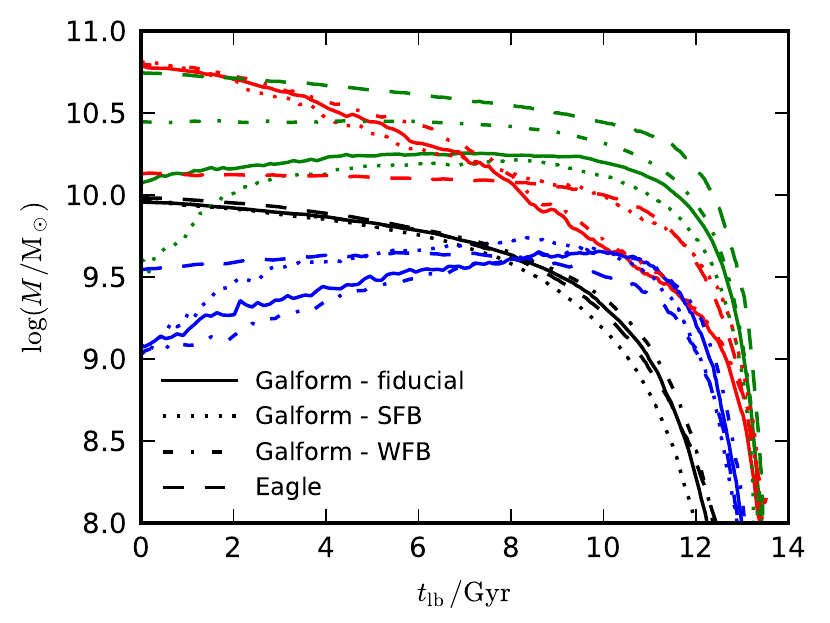}
\caption{Median baryonic mass assembly histories for central galaxies with stellar masses in the 
range, $9.75<\log(M_\star/\mathrm{M_\odot})<10.25$, at $z=0$.
Solid lines show the fiducial \galform model.
Dotted lines show a \galform variant model with strong SNe feedback (SFB) and rapid gas cycling.
Dash-dotted lines a \galform variant model with weak SNe feedback (WFB) and slow gas cycling.
Dashed lines show the \eagle reference model.
Otherwise, line formatting follows Fig.~\ref{tree_plot}.}
\label{tree_plot_model_comp}
\end{figure}

In \cite{Mitchell14} and \cite{Rodrigues17} it was demonstrated that there is a 
degeneracy between gas return timescale and mass-loading factor in \galform if
the model is calibrated to reproduce a given stellar mass function \cite[see also][]{Somerville08}. 
Intuitively, a higher mass-loading factor (more gas ejected from galaxies) can
be compensated for by increasing the rate of gas return, ultimately leading to very 
similar stellar mass growth histories. This is explicitly demonstrated in 
Fig.~\ref{tree_plot_model_comp}, which shows three \galform models. The first
is the fiducial reference model for this study (solid lines) with the return time parameter,
$\alpha_{\mathrm{return}}=1.26$ and the mass-loading factor normalization parameter
set to $V_{\mathrm{SN}} = 380 \, \mathrm{km \, s^{-1}}$. The other two models shown
were run with $\alpha_{\mathrm{return}} = 8, V_{\mathrm{SN}} = 650 \, \mathrm{km \, s^{-1}}$
(strong SNe feedback and rapid gas return, dotted lines) and 
$\alpha_{\mathrm{return}} = 0.4, V_{\mathrm{SN}} = 300 \, \mathrm{km \, s^{-1}}$
(weak SNe feedback and slow gas return, dash-dotted lines).

While the three \galform models shown predict very similar stellar mass assembly histories,
the median mass in the ejected gas reservoir ranges over almost an order of magnitude by $z=0$. Notably, the 
model with slow gas return and smaller mass-loading factors is in closer
agreement with \eagle for the evolution in mass of the ejected gas reservoir, further increasing the evidence 
that return timescales are likely longer in \eagle than are typically assumed in semi-analytic galaxy formation models.
We note also that adopting a longer return timescale alleviates the tension in our reference
model that likely unrealistic amounts of energy are implicitly injected into SNe-driven winds,
as discussed in Section~\ref{fb_subsection}.
Interestingly, \cite{Rodrigues17} show that when higher-redshift stellar mass functions 
are included as constraints, values of $\alpha_{\mathrm{return}} < 0.4$ are strongly 
disfavoured. Presumably, this tension could probably alleviated by using a scale dependent gas
return timescale, as advocated by \cite{Henriques13}.

To summarise, we have seen several indications in this section that the level of baryon
cycling after gas is blown out of galaxies by feedback is lower in \eagle than
in \galform. This supports the preliminary conclusion by \cite{Crain17} that there appears to be little
ejected gas return in \eagle. Evidence here supporting this
viewpoint comes from the significantly increased segregation in metallicity between halo
gas and ISM gas in \eagle compared to \galform and the lack of a close coevolution between
the ISM and ejected gas reservoirs in \eagle (compared to \galform). Finally, increasing
the ejected gas return timescale in \galform improves the agreement with \eagle in
baryonic mass assembly histories.

\section{Radiative cooling and infall}
\label{infall_section}

\begin{figure}
\includegraphics[width=20.5pc]{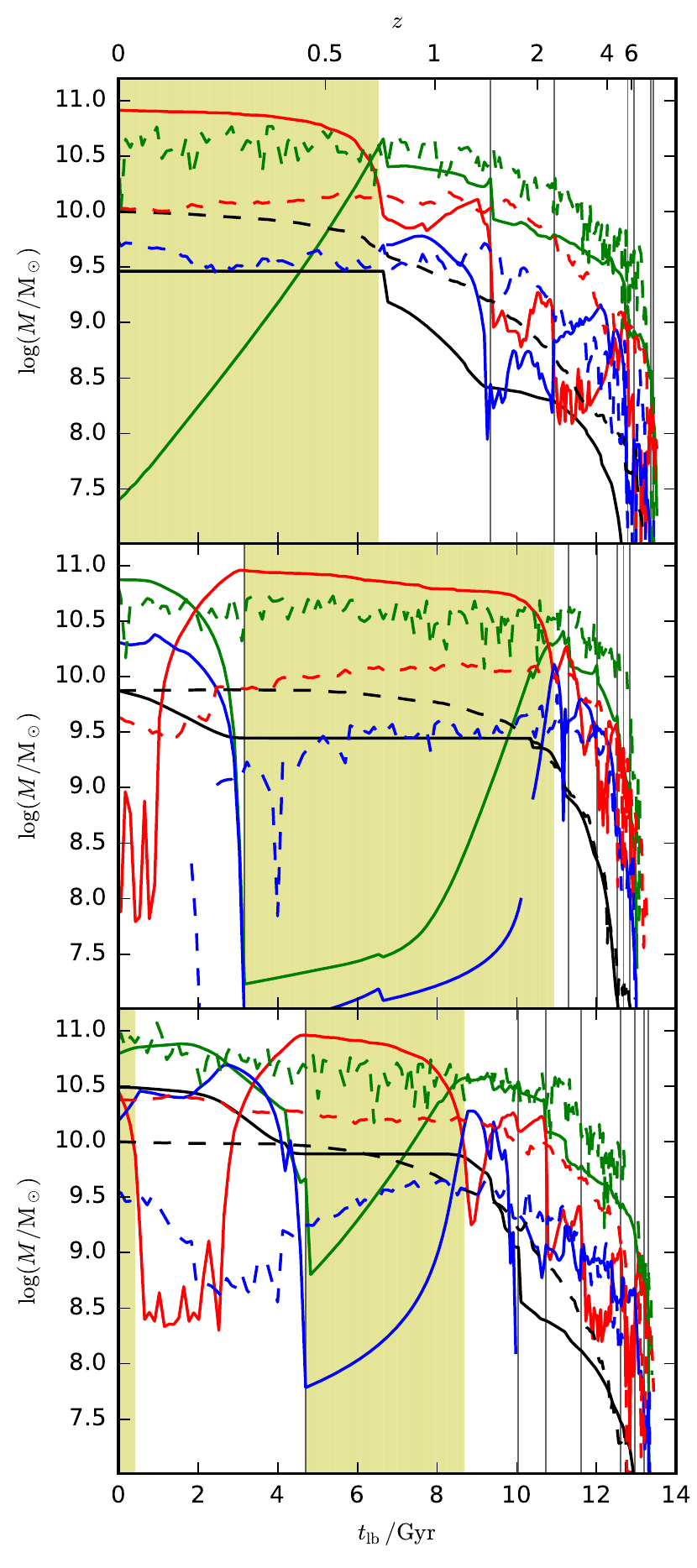}
\caption{Median baryonic mass assembly histories for examples of individual central galaxies, matched between \eagle and \galform.
Each panel corresponds to a different galaxy.
The galaxies were selected to highlight cases where \galform predicts lower, equivalent and higher stellar mass compared to \eagle in the top, middle and bottom panels respectively.
Line formatting follows \protect Fig.~\ref{tree_plot}.
Vertical grey lines show halo mass-doubling events in \galform.
Yellow shaded regions indicate when AGN actively suppress cooling from the diffuse gas halo in \galform.
}
\label{tree_plot_individ}
\end{figure}

In \galform, gas infalls from the diffuse gas halo onto the galaxy disc at a rate
which is controlled by the minimum of two physical timescales. The first is the 
radiative cooling timescale of hot halo gas and the second is the gravitational 
freefall timescale \cite[][]{Cole00,Bower06}. The limiting timescale is then compared
to the time elapsed since the previous halo mass-doubling event in order to decide how much
gas cools at the current timestep. Further details of the gas infall model are presented
in Appendix~\ref{ap:cool}.

In the regime where enough time has passed since the previous mass-doubling event for
all of the halo gas to both cool and freefall onto the galaxy disc, the gas infall
rate is simply set equal to the accretion rate of diffuse gas onto the halo.
The opposite regime occurs when AGN feedback becomes
active in a given halo, completely suppressing gas infall if the halo is considered to be quasi-hydrostatic.
The hydrostatic equality criterion used also depends on the time which has elapsed since the previous halo 
mass-doubling event \cite[][]{Bower06}.

We find that both of these regimes appear to play an active role in regulating the baryonic mass
assembly in individual haloes in \galform. This is shown directly for three example haloes
in Fig.~\ref{tree_plot_individ}. Halo mass doubling events (vertical grey lines) are followed
by a characteristic pattern. Gas rapidly infalls from the diffuse gas halo onto the disc until
either another mass-doubling event occurs or cooling abruptly stops (AGN feedback becomes active).
In some cases, there is sufficient time before one of these situations occur such that the diffuse 
gas halo is almost completely depleted (the diffuse gas halo mass
never falls to exactly zero because of details in the numerical scheme, and instead fluctuates
around a low floor value), corresponding to the regime where $r_{\mathrm{infall}} > r_{\mathrm{H}}$.

The result of this abrupt switching between regimes is the strongly oscillatory behaviour seen
for the mass in different baryonic components shown in Fig.~\ref{tree_plot_individ}. Such behaviour
is not seen for the corresponding individual haloes shown in \eagle. We deliberately select
three haloes to highlight the three possible cases for how well the stellar mass agrees between the two
models at $z=0$. In the top panel, \galform underpredicts the stellar mass at $z=0$ by a factor
$\approx 3$ compared to \eagle. In this case, the halo does not undergo a halo mass doubling event
below $z \approx 1.5$, such that AGN feedback is able to completely suppress gas infall after $z \approx 0.6$,
preventing the star formation that the corresponding \eagle galaxy undergoes during this period.
In the bottom panel, \galform overpredicts the stellar mass by the same factor compared to \eagle, in this case because
a halo mass-doubling event occurs late enough to allow significant late star formation but not so late
that AGN feedback has a significant effect at late times. In the middle panel, the two galaxies
agree in stellar mass simply because the final halo-mass doubling event in \galform fortuitously occurs
at the right time to allow the stellar mass to grow enough to match \eagle at $z=0$. 

In summary, an artificial dependence on halo mass-doubling events and a bimodal model for AGN
feedback (no effect or complete suppression of gas infall onto the disc), combined with strong SNe feedback and short return times leads to strongly oscillatory
behaviour in \galform. This behaviour is not seen in \eagle, for which it is not necessary to make 
any of these assumptions, instead allowing the associated physical phenomena to
emerge naturally (albeit with uncertain local subgrid modelling for cooling rates and energy injection from
AGN). This strongly suggests that both the cooling and AGN feedback models in \galform could be
improved with the goal of eliminating this oscillatory behaviour \cite[Hou et al., in prep, see also][]{Benson10b}.
In future work, we will directly compare the inflow rates between the two models to address
this topic more directly.

\subsection{Variable infall rates and the scatter in stellar mass}
\label{infall_scatter}

\begin{figure}
\includegraphics[width=20pc]{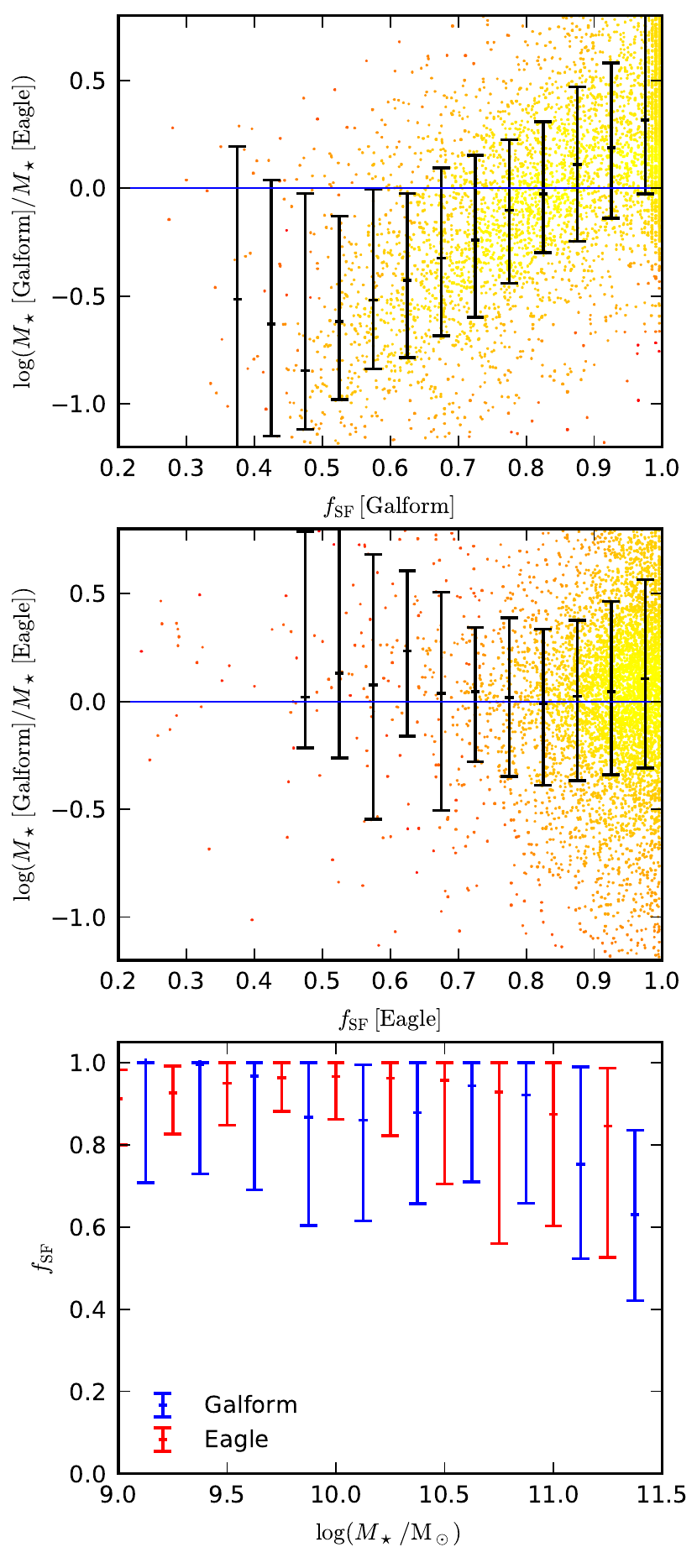}
\caption{The connection between the scatter in stellar mass for matched \galform and \eagle galaxies with star formation history.
Galaxies included are central in both \eagle and \galform at $z=0$ and have $M_\star > 10^9 \, \mathrm{M_\odot}$ in \eagle.
{\it Top:} Logarithmic residual in stellar mass at $z=0$, plotted as a function of star-forming fraction, $f_{\mathrm{SF}}$ in \galform.
Star-forming fraction is defined as the fraction of simulation outputs along the main progenitor branch for which which the main progenitor has
a specific star formation rate, $sSFR > 0.01 \, \mathrm{Gyr^{-1}}$.
Coloured points show individual galaxies.
Black points with error bars show the $16$, $50$ and $84^{\mathrm{th}}$ percentiles of the distribution.
{\it Middle:} Same but plotted as a function of star-forming fraction in \eagle.
{\it Bottom:} Star-forming fraction plotted as a function of stellar mass for \galform (blue) and \eagle (red).
In this panel, points indicate the $10$, $50$ and $90^{\mathrm{th}}$ percentiles of the distributions.
}
\label{nsf_resid}
\end{figure}

In Fig.~\ref{nsf_resid}, we attempt to relate the oscillatory behaviour in infall rates seen in
Fig.~\ref{tree_plot_individ} to the scatter in stellar mass between matched galaxies across the 
entire population shown in Fig.~\ref{stellar_mass_comp}. The top panel shows that there is a
strong positive correlation between the fraction, $f_{\mathrm{SF}}$, of simulation outputs in which a main progenitor in \galform
is considered a star-forming galaxy ($sSFR > 0.01 \, \mathrm{Gyr^{-1}}$) and the residual in stellar 
mass with the corresponding matched \eagle galaxy. Galaxies with high $f_{\mathrm{SF}}$ in \galform
form, on average, more stars than their eagle counterparts and \galform galaxies with low $f_{\mathrm{SF}}$
form fewer stars to compensate. This means a significant amount of the scatter
seen in Fig.~\ref{stellar_mass_comp} can be attributed to star formation histories being more
variable in \galform than in \eagle.

The middle panel of Fig.~\ref{nsf_resid} shows that there is not a corresponding correlation between the
stellar mass residuals and the star-forming fraction, $f_{\mathrm{SF}}$, measured in \eagle.
The lower panel of Fig.~\ref{nsf_resid} reveals that this is because the only galaxies in \eagle
with low $f_{\mathrm{SF}}$ are massive ($M_\star > 10^{10.5} \, \mathrm{M_\odot}$) and so also
have low $f_{\mathrm{SF}}$ in \galform. Conversely, there are galaxies in \galform with low $f_{\mathrm{SF}}$
at lower stellar masses ($M_\star \approx 10^{10} \, \mathrm{M_\odot}$). It is the variability
of gas infall rates in these lower mass galaxies (see Fig.~\ref{tree_plot_individ}) which are responsible for
the strong correlation seen in the top panel of Fig.~\ref{nsf_resid}.

\section{Applicability to other semi-analytic models}

Here, we briefly consider how the results from this study relate to other semi-analytic
galaxy formation models used in the community. Several of the aspects of the modelling which
we consider are specific to the details of the modelling in \galform, and cannot be easily
generalised. We consider the oscillatory infall rates tied to halo mass-doubling events and
AGN feedback (seen in Fig.~\ref{infall_scatter}) as belonging to this category. Similarly,
the effect of AGN feedback enhancing the baryon fractions of galaxy groups in \galform is 
specifically tied to the implementation of AGN feedback from \cite{Bower06}, and does not
represent the behaviour of other semi-analytic models which include an explicit Quasar-mode
of AGN feedback that can eject gas from haloes \cite[e.g.][]{Monaco07,Somerville08,Croton16}.

Of more general interest is the importance of stars forming preferentially out of gas in the
ISM with low specific angular momentum. Semi-analytic models currently assume that
disc stars form with the same specific angular momentum of either the gas disc \cite[][]{Guo11},
or the entire disc \cite[e.g.][]{Cole00,Springel05,Somerville08b,Tecce10,Croton16}.
For the specific case of the \Lgal model, cross-matching \cite{Guo16} with Appendix~\ref{ap:size_mass}
shows that this model overpredicts the sizes of low-mass 
($M_\star \approx 10^9 \, \mathrm{M_\odot}$) galaxies by a factor $\approx 2$ in the local 
Universe, consistent with \galform predictions.

Also of interest is the behaviour of gas flows both onto and out of galaxies.
For example, a wide range of gas return timescales have been adopted in contemporary
galaxy formation models. Recent models have assumed that gas returns over a halo dynamical time
\cite[][]{Lacey16}, over a Hubble time \cite[][]{Somerville08,Hirschmann16}, or have adopted more complex parametrisations where the return
time scales with halo mass or halo circular velocity \cite[][]{Guo11,Henriques13,White15,Croton16,Hirschmann16},
or where the return timescale is explicitly connected to halo growth \cite[][]{Bower12}.
\cite{Mitchell14} have also argued on empirical grounds that a yet more complex 
dependence of the gas return timescale (or SNe mass-loading factor) on halo
mass, halo dynamical time, and redshift is required to reproduce the observed
evolution of characteristic star formation rates at a given stellar mass.
With this diversity of different models, the hints found here that \eagle predicts a longer gas return
time (compared to a halo dynamical time) are certainly of interest, albeit
with the caveat that the return times are likely sensitive to the assumed heating temperature
for SNe and AGN feedback.

With respect to previous comparison studies between semi-analytic models and hydrodynamical
simulations, it is difficult to directly compare results, particularly because we have
not measured gas cooling rates in \eagle \cite[but see][]{Stevens17}.
The factor $\approx 10$ discrepancy between star formation
efficiency reported by \cite{Hirschmann12} between a set of hydrodynamical simulations 
and a semi-analytic model is not seen in Fig.~\ref{sfe} (although a $0.5 \, \mathrm{dex}$ 
discrepancy in the medians is seen for massive galaxies at $z=4$). This suggests that the 
typical star-forming gas densities of the hydrodynamical simulations analysed in \cite{Hirschmann12} 
are larger than those in \eagle, resulting in a larger disagreement with the semi-analytic model.
This may be related to the \eagle feedback model being highly effective in reducing
the typical gas densities of the ISM in which SNe explode \cite[][]{Crain15}.

\section{Summary}
\label{summary_section}

\cite{Guo16} presented a comparison between the \galform (and \Lgal) semi-analytic galaxy formation 
model and the state-of-the-art \eagle hydrodynamical simulations. They demonstrated that while the
two models are calibrated to produce similar stellar mass functions at $z=0$, the two models 
predict markedly different metallicity and galaxy size distributions as a function of stellar mass
(see also Appendix~\ref{size_mass} for a comparison of galaxy sizes).

Here, we increase the depth of the comparison by matching individual galaxies and by
isolating a number of important aspects in the physical modelling. In particular, we have carefully
assigned baryonic particles in \eagle to baryonic reservoirs that correspond to those included in \galform.
In future work, we plan to use the framework introduced here to measure the various mass, metal and
angular momentum exchanges between these reservoirs in \eagle, enabling a direct comparison to the
assumptions made in semi-analytic models pertaining to mass inflows, outflows and baryon cycling.
Even without these measurements however, a number of interesting differences between the two modelling
approaches are readily apparent at the level of detail presented here. Our main results are summarised
as follows:

\begin{itemize}

\item In Fig.~\ref{stellar_mass_comp} of Section~\ref{sm_section}, we show that the scatter in stellar mass between matched galaxies in \eagle and \galform is
$0.37 \, \mathrm{dex}$ at $z=0$ and slowly decreases with increasing redshift. For comparison,
the empirical semi-analytic galaxy formation model presented in \cite{Neistein12} achieved
an agreement with the \owls simulations of $0.08 \, \mathrm{dex}$ \cite[][]{Schaye10}. Clearly, the agreement between \galform
and \eagle could be significantly improved.

\item In Fig.~\ref{sf_threshold} of Section~\ref{threshold_section}, we show that the star formation 
thresholds implemented in \eagle and \galform lead (probably in conjunction
with differing gas surface density distributions) to strongly
differing predictions for the mass fraction of the ISM which is forming stars. This is particularly
true for low-mass galaxies at low-redshift, for which \galform predicts that almost all of the hydrogen
in the ISM is in the atomic phase and therefore not actively forming stars.

\item In Fig.~\ref{sfe} of Section~\ref{sf_eff_section}, we show that the spatially-integrated efficiency 
with which the star-forming ISM is turned into stars in \eagle
is close to a constant value at $z=0$, consistent with what is assumed in \galform. However, the local density
dependence of the Schmidt-like law implemented in \eagle leads to a star formation efficiency that increases
with redshift. \eagle also predicts that at higher redshifts the star formation efficiency is mass-dependent,
such that star formation is globally more efficient in more massive galaxies.

%\item \galform predicts higher ISM gas fractions ($M_{\mathrm{ISM}} / M_\star$) in low-mass galaxies than \eagle, in better agreement with 
%observations. This is particularly obvious at $z=0$, where the reference \eagle model predicts ISM gas fractions
%significantly below unity for $M_\star \leq 10^9 \, \mathrm{M_\odot}$.

\item \eagle predicts that the star-forming ISM typically has significantly lower specific angular momentum than
the total ISM, reflecting that it is more centrally concentrated (see Fig.~\ref{jism} in Section~\ref{j_section}). 
This is in contrast to \galform which implicitly 
assumes that star-forming gas has the same specific angular momentum as the total ISM. We show that this discrepancy could
be at least partially alleviated if \galform were to self-consistently compute the angular momentum of star-forming gas using
the radial profiles of atomic and molecular hydrogen.

\item The stellar specific angular momentum distributions as a function of stellar mass are markedly different
between \galform and \eagle (see Fig.~\ref{jstar_mstar} in Section~\ref{j_section}), 
although the interpretation is hindered because \galform does not track a specific
angular momentum for stars in galaxy bulges/spheroids. For low-mass galaxies, \galform predicts that the specific
angular momentum of galaxy discs is similar to that of their dark matter haloes (simply reflecting the assumptions of the model).
In contrast, \eagle predicts that while the ISM component of these galaxies is consistent with dark matter haloes
(at least in the higher resolution, recalibrated model), the stellar angular momentum is lower than the ISM by
$0.6 \, \mathrm{dex}$ below $z \approx 6$.

% removed for now
%\item In Fig.~\ref{vcirc} of Section~\ref{SNe_section}, we show that the choice to couple the efficiency of SNe feedback in \galform to galaxy circular velocities leads to a much higher level
%of scatter in stellar mass at a given halo mass (in the regime where there is enough mass in condensed baryons to 
%affect the gravitational potential at the halo centre) than is predicted by \eagle or inferred empirically from
%observations \cite[e.g.][]{Leauthaud12,Zu15}. This problem is compounded by the implementation of adiabatic halo contraction 
%in \galform, which serves to strengthen the coupling between stellar mass assembly and SNe feedback efficiency.

\item We show that the implementations of SNe feedback in \eagle and \galform lead to similarly low baryon fractions
in low-mass haloes ($10^{10} < M_{\mathrm{H}} \, / \mathrm{M_\odot} < 10^{11.5}$, see Fig.~\ref{fbaryon} in Section~\ref{fb_section}), 
which is the regime where AGN feedback does not play a role in the models.

\item We show \galform predicts much higher baryon fractions in group-scale haloes ($M_{\mathrm{H}} \sim 10^{13}$)
than \eagle (see Fig.~\ref{fbaryon} in Section~\ref{fb_section}). AGN feedback in \galform suppresses infall onto galaxies but does not eject gas from haloes or prevent
gas accretion onto haloes, leading to higher baryon fractions than when AGN feedback is not included. Conversely, 
including AGN feedback in \eagle acts to reduce baryon fractions in group-scale haloes.

\item We show that while the median stellar mass assembly histories of galaxies in the two models are similar, the mass
in other baryonic reservoirs is predicted to evolve differently (see Fig.~\ref{tree_plot} in Section~\ref{cycle_section}). In particular, \galform assumes that gas is
rapidly returned to the diffuse gas halo after being ejected by SNe feedback, such that the ejected gas reservoir
closely traces the evolution in the ISM reservoir. In \eagle, these reservoirs do not show such a degree of coupling in their evolution,
suggesting that the level of baryon cycling is significantly lower. Furthermore, the level of metal mixing into
the diffuse gas halo is significantly lower in \eagle than in \galform, suggesting a lower level of gas cycling.
\galform can be brought into better agreement with \eagle by using a longer gas return timescale.

\item In Fig.~\ref{tree_plot_individ} of Section~\ref{infall_section}, we show that the standard AGN and cooling models for gas infall onto galaxy discs implemented in \galform 
results in strongly oscillating infall rates for individual galaxies. This behaviour is not seen in \eagle, and it contributes
significantly to the scatter in stellar mass between matched galaxies in \galform (see Fig.~\ref{nsf_resid}). The oscillatory behaviour
in \galform stems from the implementation of AGN feedback being bimodal (either complete suppression of cooling 
or no effect) and the artificial dependency of AGN feedback and gas infall rates on halo mass doubling events.

\end{itemize}

Put together, we conclude that while galaxy evolution proceeds in a broadly similar manner 
in semi-analytic galaxy formation models compared to hydrodynamical simulations, there are 
a number of important over-simplifications adopted in \galform (but not necessarily in other 
semi-analytic models). This leads to important differences when the two models are compared 
in detail. For example, the assumption that stars form with the same specific angular momentum 
as the ISM (rather than just the star-forming ISM component) has significant consequences for 
galaxy sizes (see Appendix~\ref{ap:size_mass}).

Crucially, we have not compared inflow rates, mass-outflow rates or gas return 
timescales with \eagle in this study. We expect potentially significant differences in all
these quantities when compared to \galform and we will address this in future work. 
Specifically, it remains to be seen whether the parametrisations adopted for different physical
processes in \galform (for example, the mass loading factor associated with SNe feedback scales
as a power law with galaxy circular velocity) are capable of reproducing the macroscopic
behaviour predicted by state-of-the-art hydrodynamical simulations. Hydrodynamical 
simulations such as \eagle do not necessarily provide an accurate representation of reality.
For example, the dynamics of outflowing gas are sensitive to uncertain subgrid modelling.
Arguably however, it ought still to be possible for models
like \galform to reproduce their macroscopic behaviour once an appropriate choice of model parameters has
been adopted. That this is indeed possible has already been demonstrated
for a highly simplified model that replaces many physical considerations with a simple
empirical fit to the \owls simulations \cite[][]{Neistein12}. If a similar level of agreement
to simulations can be achieved for a more physically motivated model, semi-analytic
models can continue to be employed as useful tools for understanding galaxy evolution
with confidence that they do not make unreasonable assumptions, particularly with 
regard to angular momentum and gas cycling.

\section*{Acknowledgements}
We thank Joop Schaye for reading and providing comments that helped improve the clarity of this paper.
PM acknowledges the LABEX Lyon Institute of Origins (ANR-10-LABX-0066) of the Université de Lyon for its financial support within the program ``Investissements d'Avenir'' (ANR-11-IDEX-0007) of the French government operated by the National Research Agency (ANR).
This work was supported by the Science and Technology Facilities Council [ST/L00075X/1, ST/P000451/1].
This work used the DiRAC Data Centric system at Durham University, operated by the Institute for Computational Cosmology on behalf of the STFC DiRAC HPC Facility (www.dirac.ac.uk). 
This equipment was funded by BIS National E-infrastructure capital grant ST/K00042X/1, STFC capital grants ST/H008519/1 and ST/K00087X/1, STFC DiRAC Operations grant ST/K003267/1 and Durham University. 
DiRAC is part of the National E-Infrastructure.
CL is funded by a Discovery Early Career Researcher Award (DE150100618).

----------------------------------------------
\bibliographystyle{mn2e}
\bibliography{bibliography}
---------------------------------------------------------------------

\appendix

\section{Modelling details}
\label{ap:model_details}

\subsection{Star formation threshold}
\label{ap:sf_th}

In \eagle, a local metallicity-dependent density threshold, $n_{\mathrm{H}}^{\star}$, is used to decide which gas particles
are forming stars, given by

\begin{equation}
n_{\mathrm{H}}^{\star} = \mathrm{min}\left(0.1 \left(\frac{Z}{0.002}\right)^{-0.64},10 \right) \, \mathrm{cm^{-3}},
\end{equation}

\noindent where $Z$ is the gas metallicity \cite[][]{Schaye15}. This threshold acts to
prevent star formation taking place in diffuse and/or low metallicity gas, reflecting the physical
connection between metallicity and the formation of a cold, molecular ISM phase that can fragment
to form stars \cite[][]{Schaye04}.

In \galform, the formation of a molecular phase is explicitly
computed following empirical correlations inferred from observations \cite[][]{Lagos11a}.
Star formation is only allowed to occur in molecular gas, such that the star formation threshold
reflects the atomic/molecular ISM gas decomposition. The mass fraction of molecular hydrogen,
$R_{\mathrm{mol}} \equiv \Sigma(H_2)/\Sigma(HI)$, is computed as a function of radius
in galaxy discs by assuming a connection to the ambient pressure of the ISM in the mid-plane,
$P_{\mathrm{ext}}$, \cite[][]{Blitz06}, as

\begin{equation}
R_{\mathrm{mol}}(r) = \left(\frac{P_{\mathrm{ext}}(r)}{P_0}\right)^{0.92},
\end{equation}

\noindent where $P_{\mathrm{ext}}$ is calculated assuming vertical hydrostatic equilibrium \cite[][]{Elmegreen89}
and $P_0$ is a constant \cite[][]{Lagos11a}.
As such, the star formation threshold for discs in \galform is computed as a function of the
radial surface density profile of gas and stars and contains no metallicity dependence.
\galform also contains a distinct ISM component that represents nuclear gas that is driven into
the galaxy centre by disc instabilities and galaxy mergers. All of the gas in this component is
assumed to be in a molecular phase and is considered to be actively forming stars.

\subsection{Star formation law}
\label{ap:sf_law}

In \eagle, star-forming gas is turned into stars stochastically, sampling from a
Kennicutt-Schmidt star formation law rewritten as a pressure law by assuming
vertical hydrostatic equilibrium \cite[][]{Schaye08} such that the star formation
rate is given by

\begin{equation}
\psi = \sum_{\mathrm{i}} \, m_{\mathrm{gas,i}} \, A (1 \mathrm{M_\odot} \mathrm{pc}^{-2})^{-n} \, \left(\frac{\gamma}{G} f_{\mathrm{g}} P_{\mathrm{i}}\right)^{(n-1)/2},
\label{sfr_eqn_eagle}
\end{equation}

\noindent where $m_{\mathrm{gas,i}}$ is the gas particle mass, $\gamma=5/3$ is the ratio
of specific heats, $G$ is the gravitational constant, $f_{\mathrm{g}}$ is the gas mass
fraction (set to unity). $A$ and $n$ are treated as model parameters which are set following
direct empirical constraints from observations \cite[][]{Schaye15}. The fiducial value of
$n=1.4$ is modified to $2$ for hydrogen densities greater than $n_{\mathrm{H}} = 10^3 \, \mathrm{cm^{-3}}$.
$P_{\mathrm{i}}$ is the local gas pressure, with a pressure floor set proportional to gas density
as $P \propto \rho_{\mathrm{g}}^{4/3}$, normalized to a temperature of $T = 8 \times 10^3 \, \mathrm{K}$
at a hydrogen density of $n_{\mathrm{H}} = 0.1 \, \mathrm{cm^{-3}}$.
As such, dense star-forming gas is artificially pressurized in \eagle, ensuring that the thermal
Jeans length is always resolved, even at very high gas densities.

In \galform, the surface density of star formation in discs is linearly related to
the surface density of molecular hydrogen, following empirical constraints
from \cite{Blitz06}. The total star formation rate is therefore linearly
proportional to the star-forming ISM gas mass as

\begin{equation}
\psi = \nu_{\mathrm{SF}} M_{\mathrm{ISM,SF}},
\end{equation}

\noindent where $\nu_{\mathrm{SF}}=0.5$ is a constant empirically constrained
from observations \cite[][]{Leroy08,Bigiel11,Rahman12}. Accordingly, the efficiency of star formation
per unit star-forming ISM mass in galaxy discs is also constant.
In merger or disc-instability triggered starbursts, nuclear
gas is instead converted into stars following a decaying exponential function
\cite[][]{Lacey16}.

\subsection{Disc angular momentum}
\label{ap:j}

In \galform, disc angular momentum is computed assuming that infalling
gas from the halo conserves angular momentum \cite[][]{Cole00}. The 
total specific angular momentum of halo gas is set equal to that of the
dark matter halo, with the radial specific angular momentum profile set assuming
a constant rotation velocity. Within the disc, it is assumed that
stars and gas always have equal specific angular momentum. There
is also an assumption that disc specific angular momentum is unaffected
by stellar feedback. Bulge/spheroid angular momentum is not explicitly
modelled in \galform.

In \eagle, the angular momentum of galaxies emerges naturally from
locally solving for the laws of gravity and hydrodynamics.

\subsection{Stellar feedback}
\label{ap:sfb}

In \eagle, each star particle represents a simple stellar population with a
\cite{Chabrier03} stellar initial mass function (IMF). The gradual injection of mass
and metals through stellar evolution back into the ISM is implemented as described in
\cite{Wiersma09}. Type II SNe feedback occurs $30 \, \mathrm{Myr}$
after a stellar particle forms.
In the \galform model presented here, stars are also formed with a \cite{Chabrier03}
IMF. Unlike \eagle, \galform adopts the instantaneous recycling approximation,
whereby all of the mass and metals returned to the ISM through stellar evolution
are returned instantaneously as star formation takes place.
Correspondingly, stellar feedback is assumed to occur simultaneously with
star formation in \galform. The impact of this
assumption in semi-analytic models is addressed by \cite{Yates13}, \cite{DeLucia14}, \cite{Hirschmann16}
and Li et al., (in prep).

Stellar feedback is implemented in \eagle with the stochastic thermal energy injection scheme
of the type introduced by \cite{DallaVecchia12}. This scheme is designed to
minimise artificial radiative losses and instead allows the desired radiative losses
to be set by hand by adjusting the amount of injected thermal energy \cite[][]{Schaye15,Crain15}.
Artificial losses are effectively suppressed by requiring that neighbouring gas particles
heated by a supernova event are heated by $\Delta T = 10^{7.5} \, \mathrm{K}$, well
above the peak of the radiative cooling curve. The thermal energy injected
into the ISM per SNe is set to $f_{\mathrm{th}} \, 10^{51} \, \mathrm{ergs}$, where
$10^{51} \, \mathrm{ergs}$ is the canonical value for SNe explosions. The term $f_{\mathrm{th}}$
is parametrised as

\begin{equation}
f_{\mathrm{th}} = f_{\mathrm{th,min}} + \frac{f_{\mathrm{th,max}} - f_{\mathrm{th,min}}}{1+\left(\frac{Z}{0.1 Z_\odot}\right)^{n_{\mathrm{Z}}} \left(\frac{n_{\mathrm{H,birth}}}{n_{\mathrm{H,0}}}\right)^{-n_{\mathrm{n}}}},
\label{Eagle_eff}
\end{equation}

\noindent where $Z$ is the local gas metallicity and $n_{\mathrm{H,0}}$ is the
gas density that the stellar particle had when it formed. $f_{\mathrm{th,min}}$ and
$f_{\mathrm{th,max}}$ are model parameters that are the asymptotic values
of a sigmoid function in metallicity, with a transition scale at a characteristic
metallicity, $0.1 Z_\odot$, and with a width controlled by $n_{\mathrm{Z}}$.
An additional dependence on local gas density is controlled by model parameters,
$n_{\mathrm{H,0}}$, and $n_{\mathrm{n}}$. The two assymptotes, $f_{\mathrm{th,min}}$ and $f_{\mathrm{th,max}}$,
are set to $0.3$ and $3$ respectively. In the low metallicity, high density regime, the
energy injection therefore exceeds the canonical value for Type II SNe explosions
by a factor $3$. \cite{Crain15} and \cite{Schaye15} argue that this value is
justified on both physical and numerical grounds \cite[note also that the median
energy injection value across the simulation is lower than unity][]{Crain15}.

In \galform, rather than scale the efficiency of SNe feedback with local metallicity/gas density,
the efficiency is defined and computed globally across each galaxy disc (and separately
for galaxy bulges/spheroids). This efficiency is characterised by the dimensionless
mass loading factor, $\beta_{\mathrm{ml}}$, defined as the ratio of the mass-outflow
rate ($\dot{M}_{\mathrm{ejected}}$) from galaxies to the star formation rate ($\psi$).
Note that this is a global quantity across a given galaxy and should not be compared
to Eqn~\ref{Eagle_eff}, which pertains to the local injection of energy at a given
point in the ISM.
In \galform, $\beta_{\mathrm{ml}}$ is explicitly parametrised as a function
of galaxy circular velocity as

\begin{equation}
\beta_{\mathrm{ml}} \equiv \frac{\dot{M}_{\mathrm{ejected}}}{\psi} = \left(\frac{V_{\mathrm{circ}}}{V_{\mathrm{SN}}}\right)^{-\gamma_{\mathrm{SN}}},
\label{mass_loading}
\end{equation}

\noindent where $V_{\mathrm{SN}}$ and $\gamma_{\mathrm{SN}}$ are model parameters
that control the normalization and faint-end slope of the galaxy luminosity function
\cite[e.g.][]{Cole00} and $V_{\mathrm{circ}}$ is the galaxy circular velocity.
For star formation taking place in discs, $V_{\mathrm{circ}}$ is set to the circular
velocity of the disc at the radius enclosing half of the disc mass.
For nuclear star formation taking place in galaxy bulges/spheroids, $V_{\mathrm{circ}}$
is correspondingly set equal to the circular velocity at the half-mass radius of the
spheroid.

Unlike in \eagle, the stellar feedback in \galform does not include any energetic considerations. While the
thermal and kinetic energy of outflowing gas is not directly modelled in \galform, if we
assume (as an example) that gas is launched from galaxies in a kinetic wind with a velocity of
$250 \, \mathrm{km \, s^{-1}}$, we can estimate the galaxy circular velocity below which the energy injected exceeds the
energy available. For the SNe parameters from our fiducial model ($V_{\mathrm{SN}} = 380 \, \mathrm{km \, s^{-1}}$ and $\gamma_{\mathrm{SN}} = 3.2$),
and with a value of $8.73 \times 10^{15} \, \mathrm{erg \, g^{-1}}$ of energy available
per unit mass turned into stars \cite[as appropriate for a Chabrier IMF assuming $10^{51} \, \mathrm{erg}$ per supernova and that stars with mass $6-100 \, \mathrm{M_\odot}$ explode, ][]{Schaye15}, equating
the (example) kinetic energy of the outflowing wind with the energy available yields that
Eqn~\ref{mass_loading}
violates energetic considerations below a circular velocity,
$V_{\mathrm{circ}} = 134 \, \mathrm{km \, s^{-1}}$.
For our reference \galform model, this
circular velocity corresponds to a halo mass, $M_{\mathrm{H}} \approx 10^{12} \, \mathrm{M_\odot}$ at $z=0$.
The corresponding mass-loading factor at this velocity is very large, $\beta_{\mathrm{ml}} = 28$,
significantly in excess of the values reported by simulations at this mass scale \cite[e.g.][]{Muratov15,Christensen16,Keller16}.
The mass-loading factors predicted by \eagle will be presented in Crain et al. (in preparation) and we plan to
explicitly compare these mass loading factors with \galform in future work.

\subsection{Black hole growth and AGN feedback}
\label{ap:agn}

In \eagle, SMBH seeds are placed at the position of the highest
density gas particle within dark matter haloes of mass,
$M_{\mathrm{H}} > 10^{10} \, \mathrm{M_\odot} /h$ \cite[][]{Schaye15}. Black holes then accrete
mass with an Eddington limited, Bondi accretion rate that is modified if the accreted
gas is rotating at a velocity which is significant relative to the sound speed \cite[][]{Rosas15}.
Black holes that are sufficiently close and with sufficiently
small velocity are allowed to merge, forming a second channel of black hole growth.

Analogous to the implementation of stellar feedback, accreting SMBH particles stochastically
inject thermal energy into neighbouring gas particles. The amount of energy injected per unit
accretion contains a model parameter that controls the resulting relationship between
SMBH mass and galaxy stellar mass, but not the effectiveness of AGN feedback \cite[][]{Booth10, Schaye15, Bower17}.
This injection energy is stored in the black hole until it is sufficiently large to heat a neighbouring
gas particle by $\Delta T = 10^{8.5} \, \mathrm{K}$ which is an order of magnitude larger than the local heating
from stellar feedback ($\Delta T = 10^{7.5} \, \mathrm{K}$).

In \galform, SMBHs are seeded inside galaxies when they first
undergo a disc instability or galaxy merger event. SMBHs grow in
mass primarily by accreting a fraction of the ISM mass
converted into stars in starbursts that take place in galaxy bulges/spheroids during galaxy merger or disc
instability events \cite[][]{Bower06,Malbon07}. A second growth
channel comes from black hole mergers, which take place whenever
there is a merging event between two galaxies hosting black holes.
We compare the \galform and \eagle distributions of black hole mass
as a function of stellar mass in Appendix~\ref{ap:bh_sm}, where
we show that \galform does not predict the steep dependence on
stellar mass predicted by \eagle at $z=2,4$.

The implementation of AGN feedback in \galform is fully described in \cite{Bower06}
\cite[see also][]{Lacey16}. The most salient parts of the modeling for this analysis are
as follows.
AGN feedback in \galform is implemented such that it can be effective only when the diffuse
gas halo is in a quasi-hydrostatic state \cite[][]{Bower06}. This
occurs when the radiative cooling timescale exceeds the gravitational freefall
timescale in the diffuse gas halo. In this regime,
it is assumed that a fraction of the diffuse infalling material is directly
accreted onto the SMBH, forming a third growth channel. A fraction
of the rest mass energy of this accreted material is assumed to be injected
into the diffuse gas halo as a heating term. If this heating term
exceeds the cooling rate, infall from the diffuse gas halo onto
the ISM is assumed to be completely suppressed. As such, unlike in \eagle,
AGN feedback has no direct effect on gas in the ISM
and does not drive galactic outflows.

\subsection{Gas return timescales}
\label{ap:ret}

In \galform, gas which is ejected from galaxies is placed in a distinct
reservoir. Gas is reincorporated from this reservoir back into the diffuse 
gas halo at a rate given by

\begin{equation}
\dot{M}_{\mathrm{return}} = \alpha_{\mathrm{return}} \, \frac{M_{\mathrm{ejected}}}{\tau_{\mathrm{dyn}}},
\label{ret_time}
\end{equation}

\noindent where $\alpha_{\mathrm{return}}$ is a model parameter (typically set close to
unity and set to $1.26$ in our fiducial model), $M_{\mathrm{ejected}}$ is the mass
in the ejected reservoir and $\tau_{\mathrm{dyn}}$ is the halo dynamical time \cite[][]{Bower06}.

The spatial distribution of the ejected gas reservoir is not explicitly specified
in \galform. Whether or not the ejected gas resides within or outside
the virial radius has been subject to various interpretations as the model has
evolved over time \cite[][]{Cole00,Benson03,Bower06,Bower12}. Here, we choose
the interpretation that the ejected gas is spatially located outside the halo
virial radius for central galaxies. Physically, this corresponds to assuming
that outflowing gas leaves the virial radius over a timescale that is short
compared to other physically relevant timescales. For satellite galaxies, we consider ejected
gas to be still within the virial radius of the host halo. This interpretation allows
us to cleanly compare the indirect efficiency of feedback and the baryon cycle
with \eagle by measuring the fraction of baryons within the virial radius.

In \eagle, no explicit gas return timescale is set, as the trajectories of gas 
particles are calculated self-consistently. In practice, the return times will
be sensitive to the details of the implementations of supernova and AGN feedback,
including the heating the temperatures.

\subsection{Radiative cooling and infall}
\label{ap:cool}

In \galform, gas infalls from the diffuse gas halo onto the galaxy disc at a rate given by

\begin{equation}
\dot{M}_{\mathrm{infall}} = \frac{4 \pi \int_0^{r_{\mathrm{infall}}} \rho_{\mathrm{g}}(r) \, r^2  \,\mathrm{d}r - M_{\mathrm{cooled}}}{\Delta t}
\label{infall_eqn}
\end{equation}

\noindent where $\rho_{\mathrm{g}}(r)$ is the so-called ``notional'' gas density profile,
$M_{\mathrm{cooled}}$ is the mass that has already undergone infall from the notional gas profile
onto the disc before the current timestep and $\Delta t$ is the numerical timestep size \cite[][]{Cole00}.
$r_{\mathrm{infall}}$ is the infall radius, which represents the radius within which gas has
had sufficient time to infall from the notional profile to the disc. It is limited either by
the gravitational freefall timescale or the radiative cooling timescale.
$r_{\mathrm{infall}}$ is computed by equating the limiting radiative/freefall timescale with the time elapsed since the host halo last doubled in mass. This cooling
model was introduced in \cite{Cole00} and updated in \cite{Bower06}.

In \eagle, gas infalls onto galaxies naturally as a consequence of hydrodynamics
and gravity. Radiative cooling and photoheating are implemented element-by-element following
\cite{Wiersma09b}, assuming ionization equilibrium.

\section{Halo mass definitions}
\label{ap:halo_mass}

\begin{figure}
\includegraphics[width=20pc]{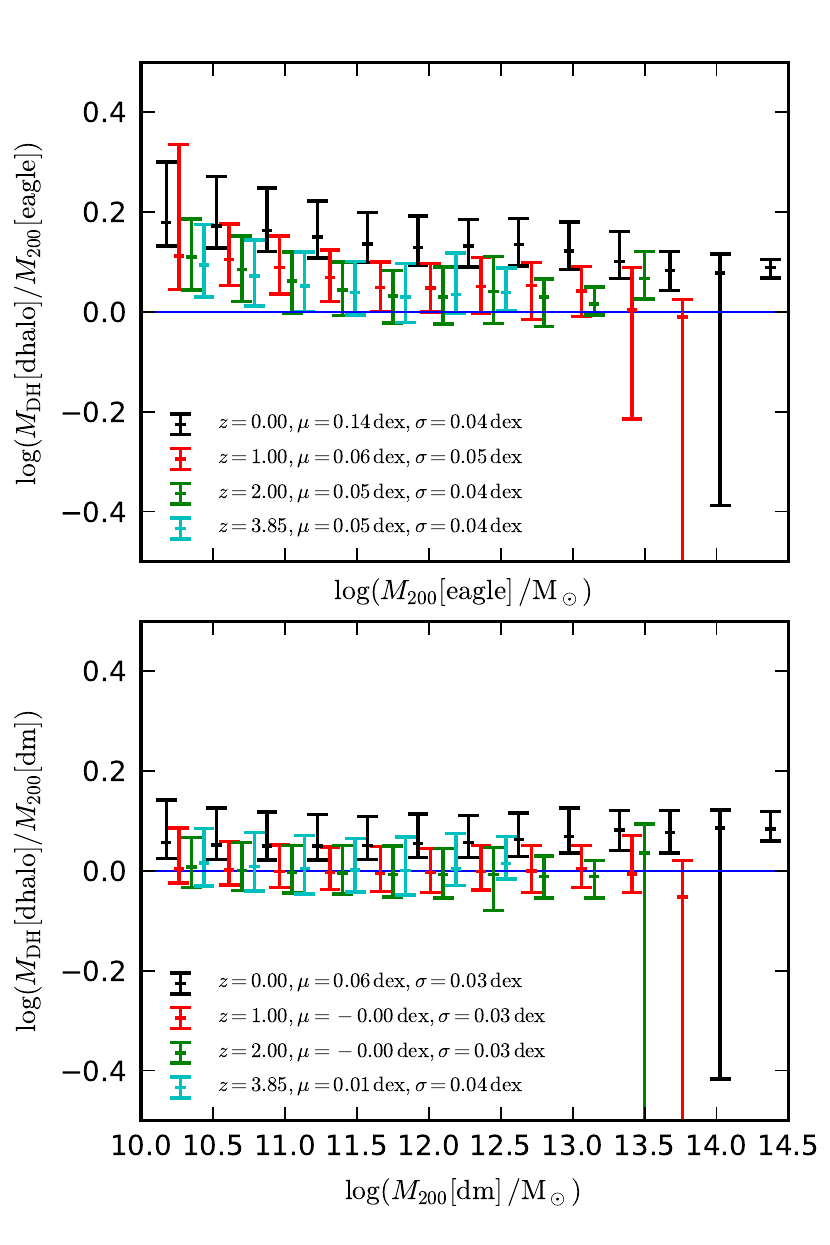}
\caption{Comparison of halo masses between different definitions and simulations.
{\it Top:} \dhalo mass used internally within \galform, $M_{\mathrm{DH}} \, \mathrm{[dhalo]}$, compared to the halo mass, $M_{\mathrm{200}} \, \mathrm{[eagle]}$, (which includes baryons) measured from the reference \eagle hydrodynamical simulation.
{\it Bottom:} \dhalo mass, $M_{\mathrm{DH}} \, \mathrm{[dhalo]}$, compared to the halo mass, $M_{\mathrm{200}} \, \mathrm{[dm]}$, measured from the \eagle dark-matter-only simulation.
Points show the $10$, $50$ and $90^{\mathrm{th}}$ percentiles of the distributions for a given redshift.
Different point colours correspond to different redshifts, as labelled.
Also labelled are the mean (logarithmic) vertical offset, $\mu$, and the mean $1 \, \sigma$ scatter.
}
\label{mhalo_comp}
\end{figure}

Fig.~\ref{mhalo_comp} compares the \dhalo halo masses used internally within \galform to the $M_{\mathrm{200}}$ halo masses 
measured from the reference hydrodynamical (top) and dark-matter-only (bottom) \eagle simulations. The difference between 
these halo mass definitions leads to a small scatter between \dhalo masses and $M_{\mathrm{200}}$ measured from the dark-matter-only simulation. There is also a 
small systematic offset at $z=0$ (this offset only appears for $z<1$) which has no trend with mass. The objects with much 
lower halo masses in \galform compared to the dark-matter-only simulation between 
$10^{13.5} < M_{\mathrm{200}} \, \mathrm{M_\odot} < 10^{14}$ are flagged as satellites by the \dhalo algorithm but are considered 
central subhaloes by \subfind, leading to the large differences between halo masses.  
Comparing \galform to the hydrodynamical simulation (top panel), the scatter is similar but with a larger, mass dependent,
offset caused primarily by the ejection of baryons by feedback in \eagle \cite[see][for a full analysis of this effect]{Schaller15b}.

\section{ISM definition in \eagle}
\label{ap:ism}

\begin{figure*}
\begin{center}
\includegraphics[width=40pc]{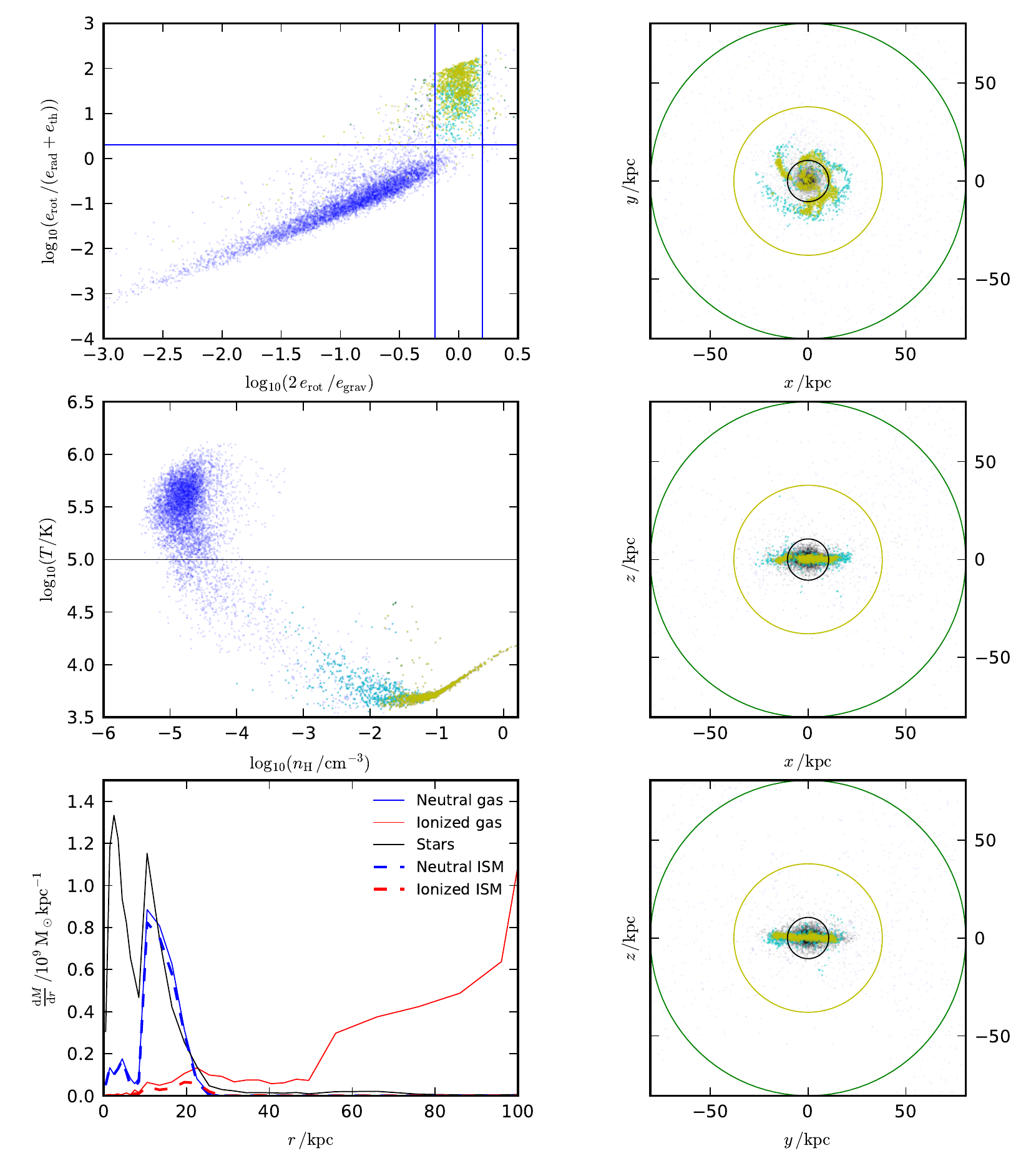}
\caption{Diagnostic information for selecting the ISM in a central, star-forming spiral galaxy with $M_\star = 10^{10} \, \mathrm{M_\odot}$ at $z=0$.
Coloured points show individual gas particles associated to the central subhalo, with yellow indicating star-forming ISM, cyan indicating non-star-forming ISM and small blue points indicating non-ISM particles.
Black points show stellar particles.
{\it Top-left:} Rotational support selection criteria. Particles above the horizontal line and between the vertical lines are considered rotational supported.
{\it Middle-left:} Phase diagram. The horizontal line shows the temperature cut, above which gas is excluded from the ISM unless it has a density, $n_{\mathrm{H}} > 50 \, \mathrm{cm^{-3}}$.
{\it Bottom-left:} Radial mass profiles of stars (black), neutral gas (neutral hydrogen and associated helium, solid blue), neutral ISM (dashed blue), ionized gas (solid red) and ionized ISM (dashed red).
{\it Right-panels:} Spatial distributions of gas and stellar particles in three projections that are face-on (top) and edge-on (middle and bottom). 
Non-ISM and stellar particles are only shown within a $80 \, \mathrm{pkpc}$ slice along the line-of-sight axis for clarity.
The black circles show twice the half-mass radius of the stellar component.
The yellow circles show twice the radius containing $90 \%$ of the pre-selected ISM mass.
Green circles show half of the halo virial radius.}
\label{spiral_z0}
\end{center}
\end{figure*}

In Section~\ref{compart_section}, we introduce selection criteria to define gas particles which belong
to an ISM component in \eagle. The canonical case demonstrating the behaviour of these selection
criteria is shown in Fig.~\ref{spiral_z0}, which shows the criteria applied to a Milky-Way-like galaxy
at $z=0$. The top-left panel demonstrates the rotational support selection criteria defined by
Eqn~\ref{rot_criteria} and Eqn~\ref{rot_criteria2}. For this galaxy, gas particles cleanly separate
into two distinct populations (the ISM and a diffuse, ionized and hot gas halo). The middle-left
panel shows the associated phase diagram, indicating that there is cool, rotationally supported ISM gas (cyan points)
at low densities which is not forming stars. The lower-left panel shows radial mass profiles, splitting
gas particles between neutral and ionized phases of hydrogen, following the methodology described in
\cite{Lagos15} and \cite{Crain17}, which utilises the self-shielding corrections from \cite{Rahmati13}.
For this galaxy, our ISM definition includes almost all of the neutral hydrogen, as well as a small amount
of amount of cool, rotationally supported, ionized hydrogen. 

The right-panels in Fig.~\ref{spiral_z0} show the spatial distribution of gas and stellar particles. 
A spiral structure for the ISM component is evident, with star-forming particles
(yellow) tracing denser regions within the spiral arms compared to non-star-forming ISM particles (cyan).
The circles in these panels relate to the various radial selection criteria described in Section~\ref{compart_section}.
The black circle indicates twice the half-mass radius of the stellar component. Dense gas within this radius
that is not considered to be rotationally supported can still be included within the ISM component. This
is why there is a small number of ISM particles (yellow/cyan) outside the selection region in the upper-left panel
of Fig.~\ref{spiral_z0}. The yellow circle in the right-panels indicates twice the radius enclosing $90 \%$ of
the mass within the ISM. Gas particles outside this radius are then excluded from the ISM. In practice, this
acts to remove a residual amount of distant, rotating material which is clearly not spatially associated to
the ISM of the central galaxy. Removing this gas has minimal impact on our results. The green circle shows
half the halo virial radius. Gas particles outside this radius are also excluded from the ISM. For this galaxy,
this radius is significantly larger than the yellow circle and so is irrelevant.

\begin{figure*}
\begin{center}
\includegraphics[width=40pc]{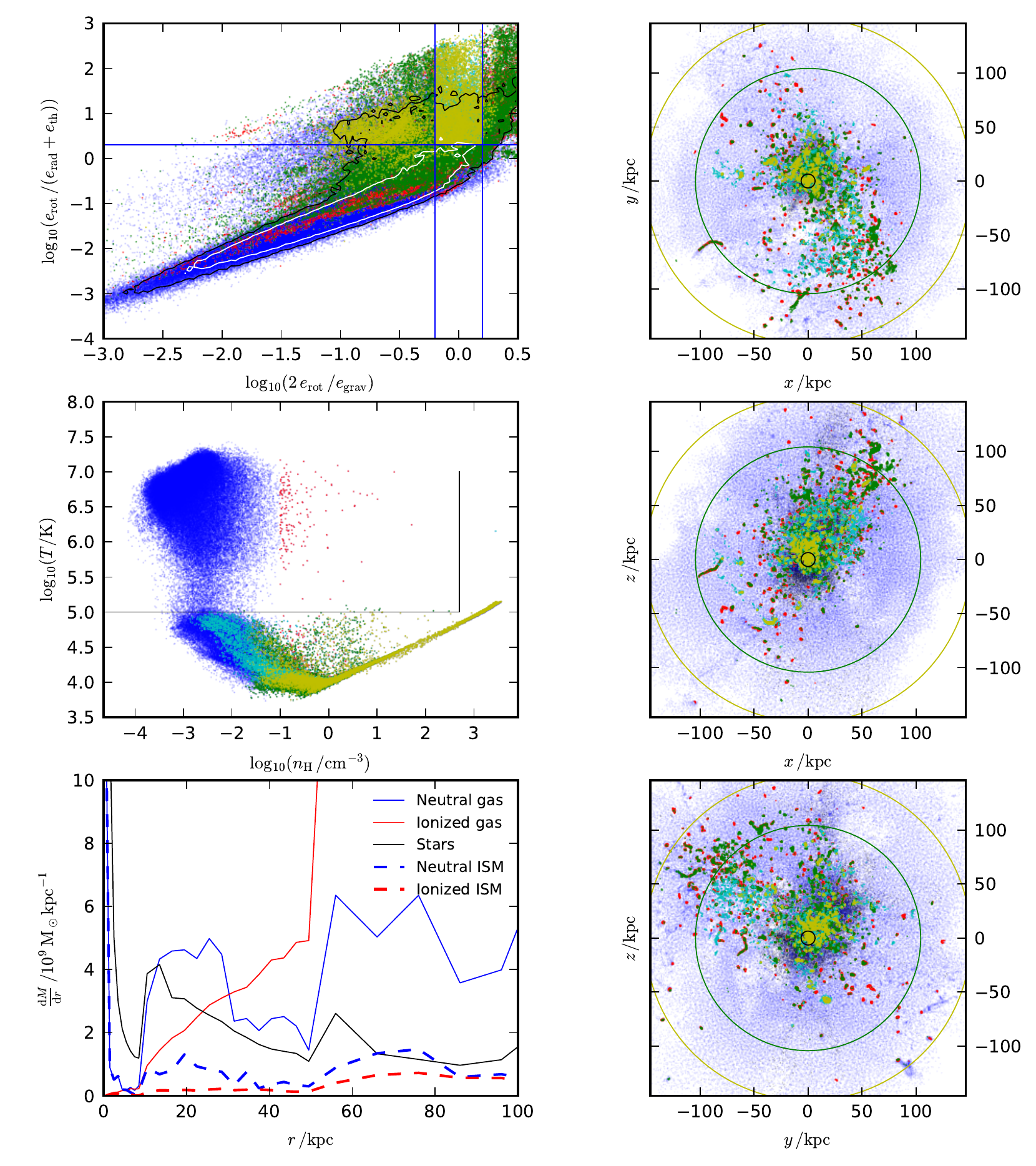}
\caption{Diagnostic information for selecting the ISM in a central, highly star-forming galaxy with $M_\star = 10^{11} \, \mathrm{M_\odot}$ at $z=2$.
Panel information as well as point and line formatting follow \protect Fig.~\ref{spiral_z0}, with the following additions.
Red points show gas particles that are dense ($n_{\mathrm{H}} > 0.1 \, \mathrm{cm^{-3}}$) but are not considered part of the ISM.
Green points (over plotted) show gas particles that are star-forming but not considered part of the ISM.
In the upper-left panel, contours are shown that enclose $90 \, \%$ (black) and $68 \, \%$ (white) of the total gas mass.}
\label{z2_monster}
\end{center}
\end{figure*}

While our ISM selection criteria appear to perform well for the galaxy shown in Fig.~\ref{spiral_z0} (and
we have checked a number of similar central/satellite examples for a variety of redshifts), other
galaxies with more extreme properties pose a greater challenge. For passive, gas-poor galaxies, we find it
is necessary to also remove gas particles from outside five times the radius enclosing half the stellar component.
In passive galaxies, there is often no clear central ISM component in the radial profiles and most of
of the cool gas is in distant, rotating clumps. Including/excluding these clumps makes little difference for
our analysis however because they form a negligible fraction of the mass in massive galaxies.

For massive, high-redshift star-forming galaxies the situation is more complex. Fig.~\ref{z2_monster} shows
the same information as Fig.~\ref{spiral_z0} but for the ``worst-case'' scenario of a massive
($M_\star = 10^{11} \, \mathrm{M_\odot}$), star-forming ($SFR = 100 \, \mathrm{M_\odot \, yr^{-1}}$) galaxy 
at $z=2$. This galaxy is an extreme example for which there is no apparent bimodality between a rotationally
supported ISM and a diffuse, hot, halo. Rather, the gas appears to be dynamically disturbed, with a very
broad mass distribution in the upper-left panel of Fig.~\ref{z2_monster}. The $1\sigma$ mass contour (white) 
encloses gas which is not considered to be in rotational equilibrium and includes a mix of hot and cool diffuse gas as well as a residual amount
of cold, dense, radially infalling gas (red and green points). The $2\sigma$ contour encloses a significant amount 
of star-forming gas which is not considered part of the ISM (green points), either
because it is rotating too slowly or too quickly or because it has too high a radial velocity to be in dynamical
equilibrium. The ISM material that is selected by the criteria described in Section~\ref{compart_section} is either rotationally
supported (and typically spatially extended) or is centrally-concentrated and is supported by a combination
of thermal pressure and rotation (yellow points to the left of the rotationally supported selection region
in the upper left panel). The relative contributions from these two components to the total ISM are roughly 
equal. Given the dynamically disturbed nature of this system, it is unclear whether the rotational equilibrium
criteria used to select the spatially extended ISM in this case are truly robust. However, the majority
of the spatially-extended neutral hydrogen which is not included in the ISM is excluded because it
is radially infalling and as such is robustly excluded.

The contrast between the situations presented in Fig.~\ref{spiral_z0} and Fig.~\ref{z2_monster} serves
to highlight the difficulty of defining the ISM across all galaxies with a uniform set of selection
criteria. Nonetheless, simply selecting star-forming gas particles would cut away $30 \, \%$ of the mass
and $40 \, \%$ of the angular momentum in the case of the well-defined ISM shown in Fig.~\ref{spiral_z0}.
Taking gas within an aperture is also likely to be overly simplistic. Too small an aperture will cut away spatially
extended, high-angular-momentum gas. Too large an aperture (even with a temperature cut) will select significant 
amounts of radially infalling gas around high-redshift galaxies that should not (at least according
to our physical criteria) be considered as part of the ISM.

\begin{figure}
\includegraphics[width=20pc]{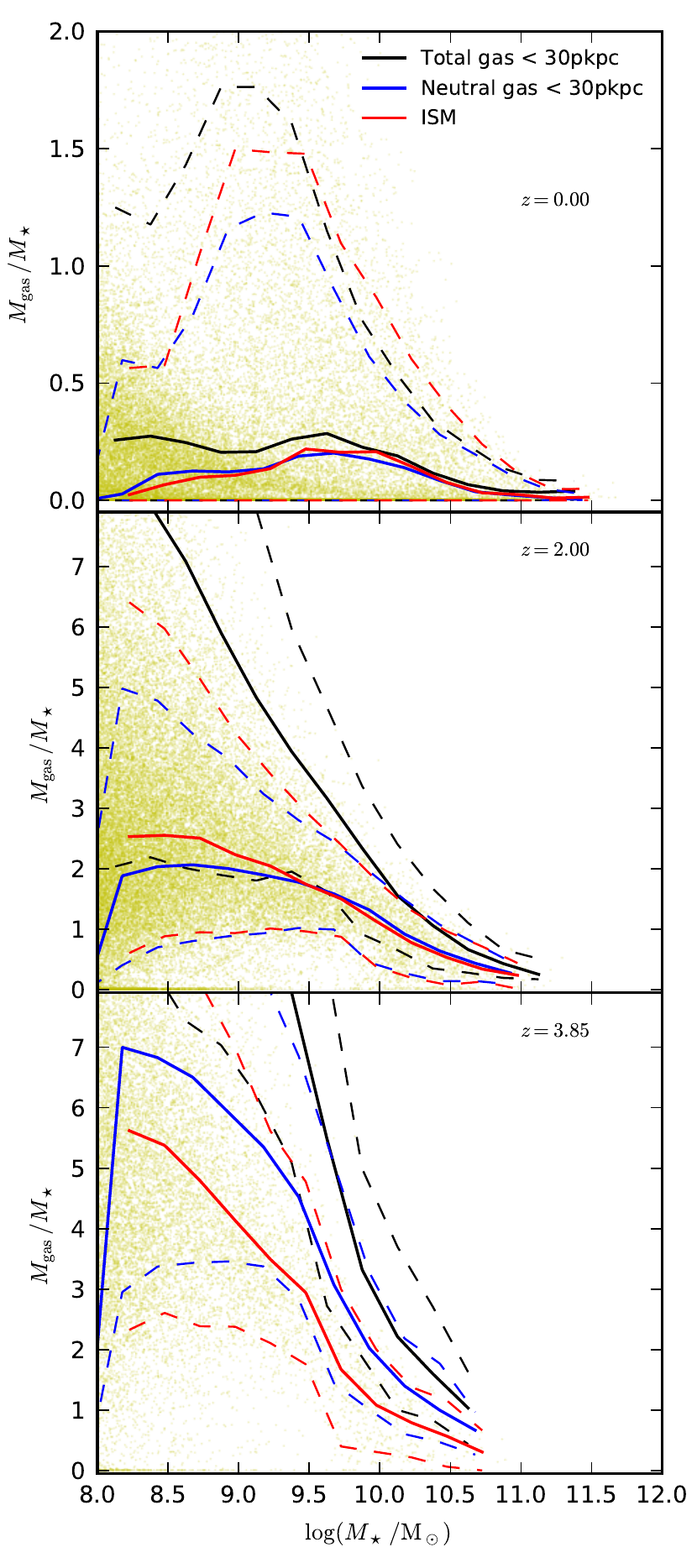}
\caption{Comparison of different possible definitions of the ISM for galaxy gas fractions.
Red lines show the distributions when using the ISM definition defined in \protect Section~\ref{compart_section}.
Blue points show the corresponding distributions when taking all neutral hydrogen (and associated helium) with a $30 \, \mathrm{pkpc}$ aperture.
Black points show the distributions when taking all gas within a $30 \, \mathrm{pkpc}$ aperture.
Solid lines show the medians and dashed lines show the $16$ and $84^{\mathrm{th}}$ percentiles of the distributions.
}
\label{gas_frac_ISM_comp}
\end{figure}

To assess the global behaviour of our criteria, Fig.~\ref{gas_frac_ISM_comp} shows the
resulting ISM gas fractions as a function of stellar mass for three redshifts. These are then
compared to the total gas within $30 \, \mathrm{pkpc}$ and the neutral hydrogen within $30 \, \mathrm{pkpc}$,
which is taken as a proxy for the ISM in \cite{Lagos15}. At $z=0$, the resulting
gas fractions are similar, indicating that the most of the hydrogen within $30 \, \mathrm{pkpc}$
of the halo centre is in a neutral phase and is in dynamical equilibrium. At $z=2$, our
ISM defintion is very close to taking neutral hydrogen within $30 \, \mathrm{pkpc}$ but
the total gas fractions (black lines) are significantly higher, presumably because of the
impact from supernova feedback in heating circumgalactic gas around high-redshift galaxies.
At $z=3.85$, our ISM definition yields systematically lower gas fractions than taking
neutral gas within an aperture, presumably by excluding neutral hydrogen present in dense,
radially infalling accretion streams.

\section{Galaxy sizes}
\label{ap:size_mass}

\begin{figure}
\includegraphics[width=20pc]{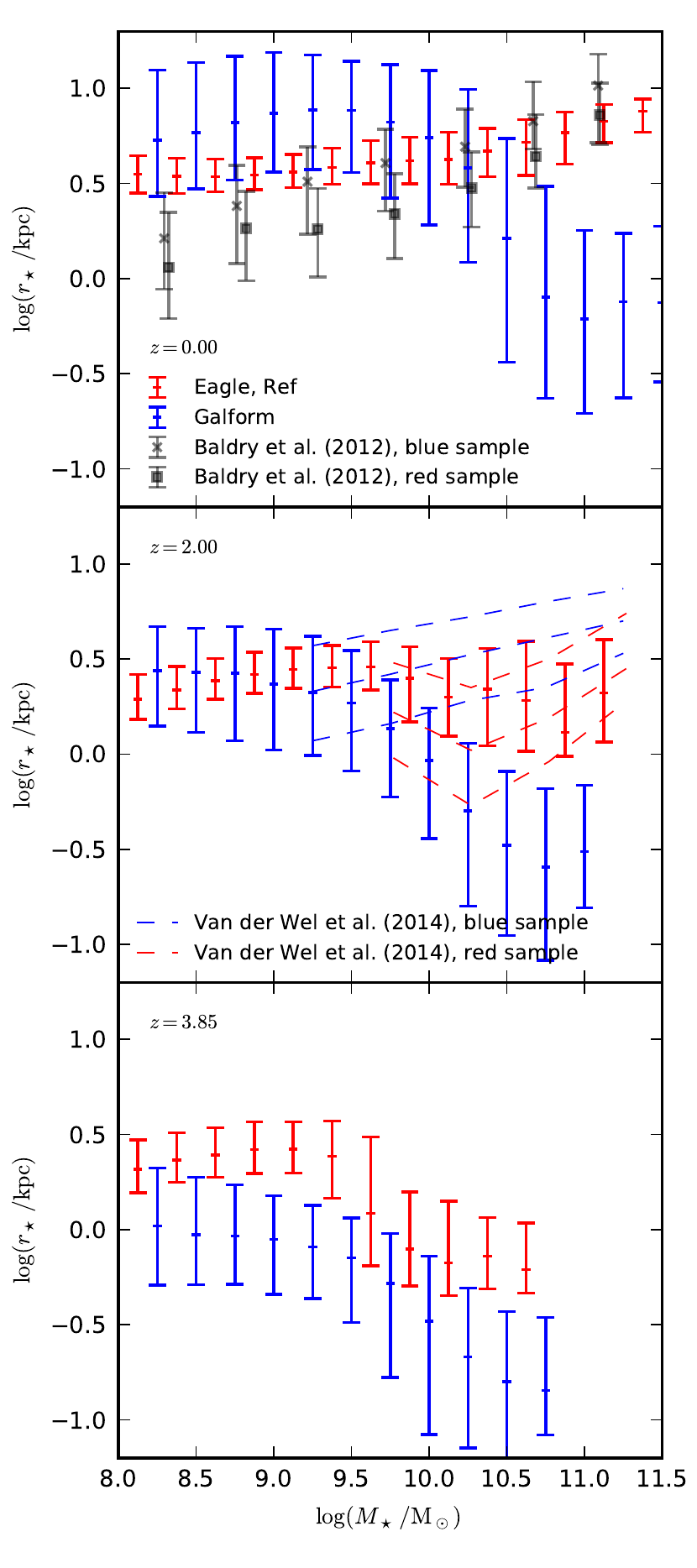}
\caption{Galaxy size distributions as a function of stellar mass.
Sizes, $r_\star$, from the reference \galform (blue) and \eagle (red) models are defined as the 3D radius enclosing half of the stellar mass (within a $30 \, \mathrm{pkpc}$ aperture for \eagle).
The points and errorbars show the $16$, $50$ and $84^{\mathrm{th}}$ percentiles of the distributions.
Grey points and errorbars show the corresponding distribution from the GAMA survey, as presented in \protect \cite{Baldry12}. Grey crosses and squares correspond to samples of blue and red galaxies respectively.
Blue and red dashed lines show respectively the distributions for blue and red galaxy samples from the CANDELS survey, as presented in \protect \cite{Vanderwel14}.
Each panel corresponds to a different redshift, as labelled.
}
\label{size_mass}
\end{figure}

Fig.~\ref{size_mass} shows the galaxy size distributions as a function of stellar mass for \eagle and \galform.
As discussed in \cite{Guo16}, the most obvious tension between the models is that \galform predicts very
compact sizes for massive galaxies. The exact underlying cause for these compact sizes is presently unclear as it
is challenging to disentangle the combined effects of modelling adiabatic halo contraction, the calculation
of pseudo-angular momentum after galaxy mergers/disc instabilities and the impact of the angular momentum histories
of progenitor galaxy discs \cite[][]{Cole00}. Any of these areas of the modelling 
could be suspect. We defer further exploration of this problem in \galform to future work.

Also apparent in Fig.~\ref{size_mass} is that the scatter in galaxy size at a fixed stellar mass is significantly
larger in \galform than in \eagle, and that the sizes of low-mass galaxies are larger in \galform than in \eagle.
We do not address the former discrepancy in this paper. The latter discrepancy is explored in Section~\ref{j_section} where 
we show that assuming that star-forming gas has the same specific angular momentum as the total ISM reservoir likely 
leads to erroneously high specific stellar angular momentum (and hence galaxy sizes) in disc-dominated (low mass, low redshift) 
galaxies.

Fig.~\ref{size_mass} also shows observational data from the GAMA survey \cite[][]{Baldry12} and the CANDELS survey 
\cite[][]{Vanderwel14}. For GAMA, two samples of red and blue galaxies are presented and the sizes quoted are effective
radii in the $i$ band. For CANDELS, two samples of star-forming and passive galaxies (determined from rest-frame colour distributions)
are presented and the sizes quoted are the semi-major axes of 1D Sersic fits at a rest-frame wavelength of $5000 \, \AA$. 
Note that we do not attempt to correct for inclination effects for sizes presented from \eagle and \galform.
While the comparison of these observed distributions to the models should be interpreted with care because of sample selection, 
projection and mass-to-light ratio effects, it is nonetheless clear that \eagle predicts a more realistic size-mass
distribution than \galform, particularly in the local Universe (where \eagle was calibrated to predict realistic sizes).

\section{Black hole masses}
\label{ap:bh_sm}

\begin{figure}
\includegraphics[width=20.01pc]{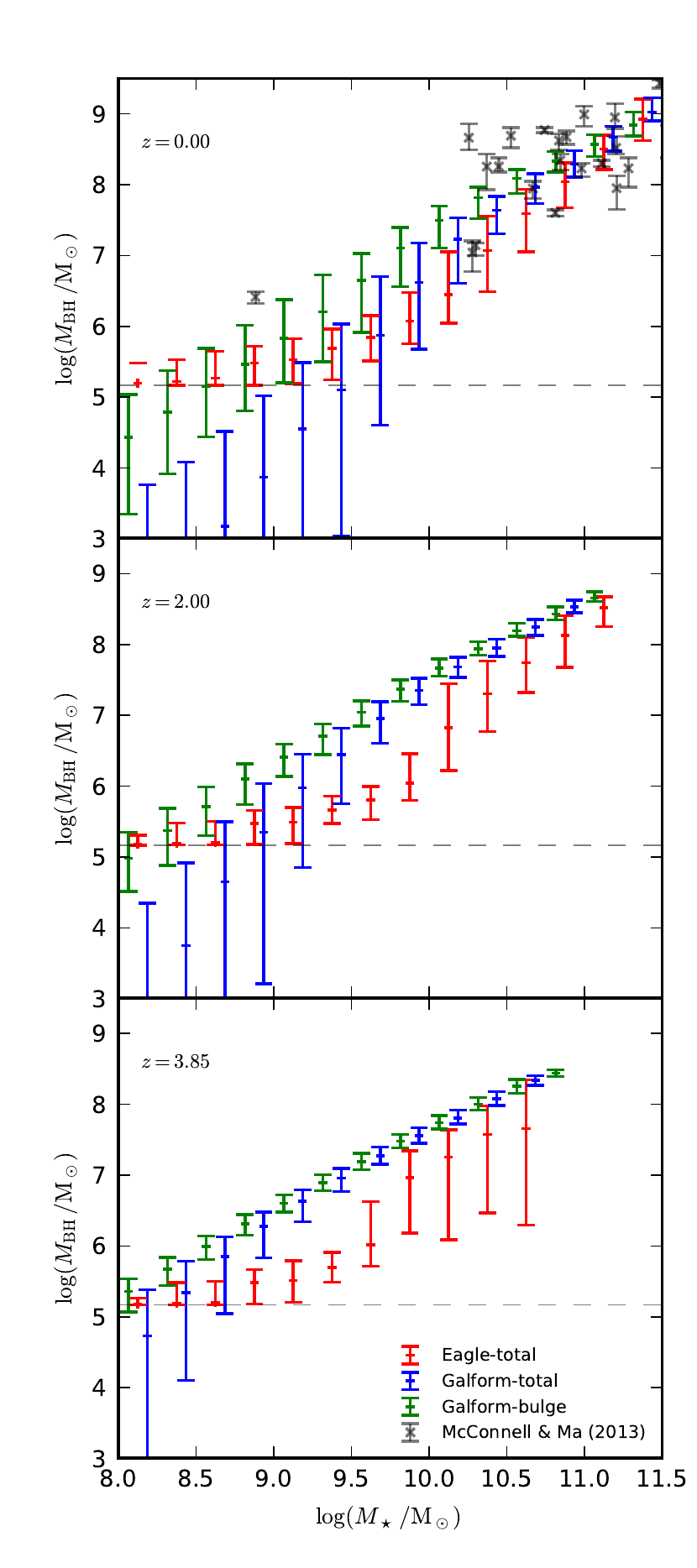}
\caption{Black hole mass as a function of stellar mass.
Red and blue points show the distribution as a function of total stellar mass from \eagle and \galform respectively.
Green points show the distribution as a function of bulge mass from \galform.
These points and errorbars show the $16$, $50$ and $84^{\mathrm{th}}$ percentiles of the distributions.
Black points show the compilation of observations from \protect \cite{McConnell13}, which are plotted as a function of bulge mass rather than total stellar mass.
Dashed grey horizontal lines indicate the black hole seed mass in \eagle.
Each panel corresponds to a different redshift, as labelled.}
\label{mbh_mstar}
\end{figure}

Fig.~\ref{mbh_mstar} shows the relationship between SMBH mass and stellar mass in the two models.
At $z=0$, the two models are very similar for $M_\star > 10^{10} \, \mathrm{M_\odot}$. For
$9<\log(M_\star \, / \mathrm{M_\odot})<10$, \galform predicts a significantly larger scatter in
SMBH mass. Unlike in \eagle, in \galform black hole growth is explicitly coupled to the
growth of the galaxy bulge. The large scatter therefore reflects the
significant scatter in bulge-to-total stellar mass ratio predicted by \galform in this mass range.
The fraction of bulge stars that were formed quiescently in progenitor discs (versus bulge stars that
were formed in galaxy merger or disc instability triggered star bursts) also plays a role in shaping the scatter in SMBH mass.
At lower masses, \eagle is affected by the seed mass, rendering a comparison meaningless.

Interestingly, at higher redshifts \eagle predicts lower black hole masses compared to \galform,
and a much steeper dependence with stellar mass at high masses. This effect is discussed extensively in
\cite{Bower17}, who interpret SMBH growth in \eagle as governed by a strongly non-linear
transition in SMBH accretion efficiency that occurs at a characteristic halo mass scale. This
scale is associated with the scale at which a hot corona develops, preventing SNe from
driving a buoyant outflow. By construction, a strongly non-linear accretion efficiency transition does not
emerge in \galform, leading to a shallower SMBH-stellar mass relation.

\label{lastpage}
\end{document}